\documentstyle[12pt]{article}

\newcommand{\be}{\begin{equation}}
\newcommand{\ee}{\end{equation}}
\newcommand{\Dlt}{\Delta}
\newcommand{\dlt}{\delta}

\newcommand{\vp}{\varphi}
\newcommand{\ep}{\varepsilon}

\newcommand{\cD}{{\cal D}}

\newcommand{\cL}{{\cal L}}

\newcommand{\cM}{{\cal M}}

\begin{document}

\begin{center}
{\Large{\bf Decision Theory with Prospect Interference and
Entanglement} \\ [5mm]

V.I. Yukalov$^{1,2}$ and D. Sornette$^{1,3}$
} \\ [5mm]

{\it $^1$Department of Management, Technology and Economics, \\
ETH Z\"urich, Z\"urich CH-8032, Switzerland,\\ [2mm]
$^2$Bogolubov Laboratory of Theoretical Physics, \\
Joint Institute for Nuclear Research, Dubna 141980, Russia \\ [2mm]
$^3$Swiss Finance Institute,\\
c/o University of Geneva, 40 blvd. Du Pont d'Arve, CH 1211 Geneva 4,
Switzerland}

\end{center}

\vskip 5cm

{\parindent=0pt
{\bf Address for correspondence}:

\vskip 2mm

Prof. V.I. Yukalov \\
Department of Management, Technology and Economics, \\
ETH Z\"urich, Z\"urich CH-8032, Switzerland

\vskip 2mm
{\bf Office phone}: +41 (44) 632 - 0848

{\bf E-mail}: yukalov@theor.jinr.ru }

\newpage

\begin{abstract}

We present a novel variant of decision making based on the 
mathematical theory of separable Hilbert spaces. This mathematical 
structure captures the effect of superposition of composite prospects, 
including many incorporated intentions, which allows us to describe a 
variety of interesting fallacies and anomalies that have been 
reported to particularize the decision making of real human beings. 
The theory characterizes entangled decision making, non-commutativity 
of subsequent decisions, and intention interference. We demonstrate 
how the violation of the Savage's sure-thing principle, known as   
the disjunction effect, can be explained {\it quantitatively} as a result 
of the interference of intentions, when making decisions under 
uncertainty. The disjunction effects, observed in experiments, are 
accurately predicted using a theorem on interference alternation that we 
derive, which connects aversion-to-uncertainty to the appearance of 
negative interference terms suppressing the probability of actions. The 
conjunction fallacy is also explained by the presence of the interference 
terms. A series of experiments are analysed and shown to be in 
excellent agreement with a priori evaluation of interference effects. 
The conjunction fallacy is also shown to be a sufficient condition for 
the disjunction effect and novel experiments testing the combined 
interplay between the two effects are suggested.

\end{abstract}

\vskip 1cm


\vskip 1cm

JEL CLASSIFICATIONS: C10, C40, C44, D03

\newpage

\section{Introduction}

Decision theory is concerned with identifying what are the 
optimal decisions and how to reach them. Most of decision theory 
is normative and prescriptive, and assumes that people are 
fully-informed and rational. These assumptions have been 
questioned early on with the evidence provided by the Allais 
paradox (Allais, 1953) and many other behavioral paradoxes 
(Camerer et al., 2003), showing that humans often seem to deviate 
from the prescription of rational decision theory due to cognitive 
and emotion biases. The theories of bounded rationality (Simon, 
1955) of behavioral economics and of behavioral finance have 
attempted to account for these deviations. As reviewed by Machina 
(2008), alternative models of preferences over objectively or 
subjectively uncertain prospects have attempted to accommodate 
these systematic departures from the expected utility model while 
retaining as much of its analytical power as possible. In particular,
non-additive nonlinear probability models have been developed to 
account for the deviations from objective to subjective probabilities 
observed in human agents (Quiggin, 1982; Gilboa, 1987; Schmeidler, 
1989; Gilboa and Schmeidler, 1989; Cohen and Tallon, 2000; 
Montesano, 2008). However, many paradoxes remain unexplained 
or are sometimes rationalized on an ad hoc basis, which does not 
provide much predictive power. Various attempts to extend
utility theory by constructing non-expected utility functionals
(Machina, 2008) cannot resolve the known classical paradoxes (Safra
and Segal, 2008). Moreover, extending the classical utility theory 
"ends up creating more paradoxes and inconsistencies than it resolves" 
(Al-Najjar and Weinstein, 2009).  

Here, we propose a novel approach, developed as a part of the 
mathematical theory of Hilbert spaces (Dieudonn\'e, 2006) and 
employing the mathematical techniques that are used in quantum theory. 
Because of the latter, we call this approach the {\it Quantum Decision 
Theory} (QDT). This approach can be thought of as the mathematically 
simplest and most natural extension of objective probabilities into 
nonlinear subjective probabilities. The proposed formalism allows us 
to explain {\it quantitatively} the disjunction and conjunction effects. 
The disjunction effect is the failure of humans to obey the sure-thing 
principle of classical probability theory. The conjunction effect  
is a logical fallacy that occurs when people assume that specific 
conditions are more probable than a single general one. Our QDT 
unearths a deep relationship between the conjunction and the 
disjunction effects. We show that the former is sufficient for the 
later to exist.

QDT uses the same underlying mathematical structure as the one
developed to establish a rigorous formulation of quantum mechanics
(von Neumann, 1955). Based on the mathematical theory of separable 
Hilbert spaces, quantum mechanics showed how to reconcile and combine 
the continuous wave description with the fact that waves are organized 
in discrete energy packets, called quanta, that behave in a manner 
similar to particles. Analogously, in our framework, the qualifier 
quantum emphasizes the fact that a decision is a discrete selection 
from a large set of entangled options. Our key idea is to provide the 
simplest generalization of the classical probability theory underlying 
decision theory, so as to account for the complex dynamics of the many 
nonlocal hidden variables that may be involved in the cognitive and 
decision making processes of the brain. The mathematical theory of 
complex separable Hilbert spaces provides the simplest direct way to 
avoid dealing with the unknown hidden variables, and at the same time 
reflecting the complexity of nature (Yukalov, 1975). In decision making, 
the hidden variables can be the many unknown states of nature, the 
emotions, and subconscious processes.

However, it is necessary to stress that our approach does not require 
that a decision maker be a quantum object. All analogies with quantum 
processes have to be understood solely as mathematical analogies helping
the reader to catch why the functional analysis is really an appropriate 
tool for modeling decision making. Before presenting our approach, it 
is useful to briefly mention previous studies of decision making and of 
the associated cognitive processes of the brain which, superficially, 
could be considered as related to our approach. This exposition will 
allow us to underline the originality and uniqueness of our approach. 
We do not touch here purely physiological aspects of the problem, which 
are studied in medicine and the cognitive sciences. Concerning the 
functional aspects of decision making, we focus our efforts towards its 
{\it mathematical modeling}.

Two main classes of theories invoke the qualifier ``quantum".
In the first class, one finds investigations which attempt to represent
the brain as a quantum or quantum-like object (Penrose, 1989; Lockwood,
1989; Satinover, 2001), for which several mechanisms have been suggested
(Fr\"ohlich, 1968; Stuart et al., 1978, 1979; Beck and Eccles, 1992;
Vitiello, 1995; Hagan et al., 2002; Pessa and Vitiello, 2003). The
existence of genuine quantum effects and the operation of any of these
mechanisms in the brain remain however controversial and have been
criticized by Tegmark (2000) as being unrealistic. Another approach 
in this first class appeals to the mind-matter duality, treating
mind and matter as complementary aspects and considering consciousness
as a separate fundamental entity (Chalmers, 1996; Atmanspacher et al.,
2002; Primas, 2003; Atmanspacher, 2003). This allows one, without
insisting on the quantum nature of the brain processes, if any, to 
ascribe quantum properties solely to the consciousness itself, as has 
been advocated by Stapp (1993, 1999). 

Actually, the basic idea that mental processes are similar to 
quantum-mechanical phenomena goes back to Niels Bohr. One of the first 
publications on this analogy is his paper (Bohr, 1929). Later on, he 
discussed many times the similarity between quantum mechanics and the 
function of the brain, for instance in Bohr (1933, 1937, 1961). This 
analogy proposes that mental processes could be modeled by 
quantum-mechanical wave functions, with all the consequences following 
from the mathematical properties of these objects. One of such immediate 
consequences would be the appearance of interference effects that are 
typical of quantum mechanics.

The second class of theories does not necessarily assume quantum properties
of the brain or that consciousness is a separate entity with quantum
characteristics. Rather, these approaches use quantum techniques, as a
convenient language to generalize classical probability theory. An example
is provided by so-called quantum games (Meyer, 1999; Goldenberg et al.,
1999; Eisert and Wilkens, 2000; Johnson, 2001; Benjamin and Hayden 2001;
Iqbal and Toor, 2001; Li et al., 2001; Du et al., 2001, 2002; Lee
and Johnson, 2003). According to van Enk and Pike (2002), any quantum 
game can be reformulated as a classical game rigged with some additional 
conditions. Another example is the Shor (1997) algorithm, which is purely 
quantum-mechanical, but is solving the classical factoring problem. This 
shows that there is no contradiction in using quantum techniques for 
describing classical problems. 

In any case, whether we deal really with a genuine quantum system or 
with an extremely complex classical system, the language of quantum theory 
can be a convenient effective tool for describing such complex systems 
(Yukalov, 1975). When dealing with genuinely quantum systems, the QDT 
provides natural algorithms that could be used for quantum information 
processing, the operation of quantum computers, and in creating artificial 
quantum intelligence (Yukalov and Sornette, 2008, 2009). In the case of 
decision making performed by real people, the subconscious activity and 
the underlying emotions, which are difficult to quantify, play the role 
of the hidden variables appearing in quantum theory. 

It is important to stress that we do not assume that human brain has 
anything to do with a real quantum object or that consciousness possesses 
some underlying quantum nature. But we use the theory of complex 
separable Hilbert spaces as a {\it mathematical language} that is 
convenient for the formal description of complicated processes associated 
with decision making. What we actually need is just the mathematical 
theory of Hilbert spaces. We could even avoid the use of the term 
"quantum", since there is no any quantum mechanics, as a physical theory,
in our approach. The sole common thing between our QDT and quantum 
mechanics is that both employ the theory of 
Hilbert spaces, characterizing the states as vectors in this space. We 
use the denomination "quantum" for brevity and because quantum theory
is also based on the theory of Hilbert spaces. In that way, we employ
the techniques of quantum theory as a convenient mathematical tool, 
without assuming any genuine underlying quantum processes. 

As another analogy, we can mention the theory of differential equations, 
which was initially developed for describing the motion of planets. But 
later on, this theory has been extended to numerous problems, having 
nothing to do with the motion of planets, and employed in a variety of 
branches of science as a mathematical tool. To emphasize this point, we 
conclude the section by the important statement that clarifies our 
position and helps the reader avoid any confusion.   

\vskip3mm

{\bf Statement}. {\it Quantum Decision Theory is based on the 
mathematical techniques employed in quantum theory, using the notion of 
Hilbert spaces as a formal mathematical tool. But QDT does not require 
that a decision maker be necessarily a quantum object}.

\section{Foundations of Quantum Decision Theory}

The classical approaches to decision making are based on the utility
theory (von Neumann and Morgenstern, 1944; Savage, 1954). Decision making
in the presence of uncertainty about the states of nature is formalized
in the statistical decision theory (Lindgren, 1971; White, 1976; Hastings
and Mello, 1978; Rivett, 1980; Buchanan, 1982; Berger, 1985; Marshall and
Oliver, 1995; Bather, 2000; French and Insua, 2000; Raiffa and Schlaifer,
2000; Weirich, 2001). Some paradoxes, occurring in the interpretation of
classical utility theory and its application to real human decision
processes have been discussed, e.g., by Berger (1985), Zeckhauser (2006), 
and Machina (2008).

\subsection{Idea of Quantum Decision Theory}

Here we suggest another approach to decision making, which is principally
different from the classical utility theory. We propose to define the
action probability as is done in quantum mechanics, using the mathematical 
theory of complex separable Hilbert spaces. This proposition can be 
justified by invoking the following analogy. The probabilistic features 
of quantum theory can be interpreted as being due to the existence
of the so-called nonlocal hidden variables. The dynamical laws of these
nonlocal hidden variables could be not merely extremely cumbersome, but
even not known at all, similarly to the unspecified states of nature. The
formalism of quantum theory is then formulated in such a way as to avoid
dealing with unknown hidden variables, but at the same time to reflect the
complexity of nature (Yukalov, 1975). In decision making, the role of hidden
variables is played by unknown states of nature, by emotions, and by
subconscious  processes, for which quantitative measures are not readily
available.

In the following sub-sections, we develop the detailed description of
the suggested program, explicitly constructing the action probability in
quantum-mechanical terms. The probability of an action is intrinsically
subjective, as it must characterize intended actions by human beings. For
brevity, an intended action can be called an {\it intention} or an 
{\it action}. In compliance with the terminology used in the theories of 
decision-making, a composite set of intended actions, consisting of several 
subactions, will be called a {\it prospect}. An important feature of our 
approach is that we insist on the necessity of dealing not with separate 
intended actions, but with composite prospects, including many 
incorporated intentions. Only then it becomes possible, within the frame of 
one general theory, to describe a variety of interesting unusual phenomena 
that have been reported to characterize the decision making properties 
of real human beings.

Mathematically, our approach is based on the von Neumann theory of quantum 
measurements (von Neumann, 1955). The relation of the von Neumann theory 
to quantum communication procedures has been considered by Benioff (1972). 
We generalize the theory to be applicable not merely to simple actions, 
but also to composite prospects, which is of principal importance for the 
appearance of decision interference. A brief account of the axiomatics of 
our approach has been published in the recent letters (Yukalov and Sornette, 
2008, 2009). The aim of the present paper is to provide a detailed 
explanation of the theory and to demonstrate that it can be successfully 
applied to the real-life problems of decision making.  

\subsection{Main Definitions}

In order to formulate in precise mathematical terms the process of
decision making, it is necessary  to introduce several definitions.
To better understand these definitions, we shall give some very simple
examples. The entity concerned with the decision making task can be a 
single human, a group of humans, a society, a computer, or any other 
system that is able or enables to make decisions. Throughout the paper, 
we shall employ the Dirac (1958) notations widely used in quantum theory. 

\vskip 3mm
{\bf Definition 1. Intended actions}

\vskip 2mm
An intended action which, for brevity, can be called an {\it intention} 
or an {\it action}, is a particular thought about doing something. 
Examples of intentions could be as follows: ``I would like to marry" or 
``I would like to be rich" or ``I would like to establish a firm". There 
can be a variety of intentions $A_i$, which are enumerated by the 
index $i=1,2,3,\ldots$. Between any two intended actions, $A$ and $B$, 
it is possible to define the binary operations of addition and 
multiplication in the same way as it is standardly done in mathematical 
logic (Mendelson, 1965) and probability theory (Feller, 1970). The sum
$A+B$ means that either $A$ or $B$ is intended to be accomplished. The
summation of several actions is denoted as 
$\bigcup_i A_i \equiv A_1 + A_2 + \cdots$. The product $AB$ implies that
both $A$ and $B$ are intended to be accomplished together. The product 
of several intended actions is denoted as 
$\bigcap_i A_i \equiv A_1 A_2 \cdots$. The total set of such intended 
actions, equipped with these binary operations, is called the {\it action 
ring}.  

\vskip 3mm
{\bf Definition 2. Action modes}

\vskip 2mm
Intention representations, or action modes, are concrete implementations 
of an intention. For instance, the intention ``to marry" can have as 
representations the following variants: ``to marry $A$" or ``to marry $B$", 
and so on. The intention ``to be rich" can have as representations ``to be 
rich by working hard" or ``to be rich by becoming a bandit". The intention  
``to establish a firm" can have as representations ``to establish a firm 
producing cars" or ``to establish a firm publishing books" and so on. We 
number the representations of an $i$-intention by the index 
$\mu=1,2,3,\ldots, M_i$. The intention representations may include not only 
positive intention variants ``to do something" but also negative variants 
such as ``not to do something". For example, the Hamlet's hesitation 
``to be or not to be" is the intention consisting of two representations, 
one positive and the other negative.

\vskip 3mm
{\bf Definition 3. Mode states}

\vskip 2mm
The mode state, or representation state, of an action mode $A_{i\mu}$ is 
denoted as the vector $|A_{i\mu}>$ corresponding to the $\mu$-representation 
of an $i$-intention. This vector is a member of a linear space to be 
defined below.

\vskip 3mm
{\bf Definition 4. Mode basis}

\vskip 2mm
The mode basis, or representation basis, is the set $\{|A_{i\mu}>\}$ 
of the representation states $|A_{i\mu}>$ corresponding to those 
intention representations $A_{i\mu}$, which are classified as basic. Here 
``basic" means the most important and fundamental, in the sense that  
linear combinations of the vectors $|A_{i\mu}>$ exhaust the whole set of 
$i$-intentions. The members of a mode basis are supposed to be well 
distinguished from each other and also normalized. This can be formalized 
as saying that the representation basis is orthonormal, which implies that 
a form, called scalar product, is defined, such that the scalar product 
$<A_{i\mu} | A_{i\nu}>$ yields the Kronecker delta symbol $\dlt_{\mu\nu}$:
\be
\label{1}
<A_{i\mu} | A_{i\nu}> = \dlt_{\mu\nu} \;  .
\ee

\vskip 3mm
{\bf Definition 5. Mode space}

\vskip 2mm
The mode space consists of all possible intention states. It is formed as 
the closed linear envelope
\be
\label{2}
\cM_i \equiv \overline \cL \{ |A_{i\mu}> \} \;
\ee
spanning the mode basis $\{|A_{i\mu}>\}$. Thus, we can assume that the 
mode space is a Hilbert space, that is, a complete normed space, with 
the norm generated by the scalar product.

\vskip 3mm
{\bf Definition 6. Intention states}

\vskip 2mm
The intention state at time $t$ is a function 
\be 
\label{3}
|\psi_i(t)> = \sum_{\mu} c_{i\mu}(t)| A_{i\mu}> \; , 
\ee
corresponding to an $i$-intention, which can be represented as a linear 
combination of the states from the representation basis $\{|A_{i\mu}>\}$. 
The intention state (\ref{3}) is a member of the mode space (\ref{2}).
Since the mode space has been assumed to be a Hilbert space, the 
associated scalar product exists and yields 
\be 
\label{4} 
<\psi_i(t_1)| \psi_i(t_2) > \;  \equiv \sum_{\mu} c_{i\mu}^*(t_1)
c_{i\mu}(t_2)  \; . 
\ee 
The norm of the intention state (\ref{3}) is generated by the scalar 
product (\ref{4}) as 
\be 
\label{5}
|| |\psi_i(t)>|| \equiv \sqrt{ < \psi_i(t) | \psi_i(t)> } \; . 
\ee 
The expansion coefficients in Eq. (\ref{3}) are assumed to be
defined by the decision maker, so that $|c_{i\mu}|^2$ gives the
weight of the state $|A_{i\mu}>$ into the general intention state.

\vskip 3mm
{\bf Definition 7. Action prospects}

\vskip 2mm
A prospect $\pi_j$ is a conjunction of several intended actions or several 
intention representations. In reality, an individual is always motivated 
by a variety of intentions, which are mutually interconnected. Even the 
realization of a single intention always involves taking into account 
many other related intentions. So, generally, a prospect is an object of 
the composite type $ABC\cdots$, where each action can be composed of 
several modes.

\vskip 3mm
{\bf Definition 8. Elementary prospects}

\vskip 2mm
An elementary prospect $e_n$ is a simple prospect formed by a conjunction 
of single action modes $A_{i\nu_i}$. With each intention representation 
marked by the index $\nu_i$, the elementary prospect is labelled by the 
multi-index
\be
\label{6}
n \equiv \{ \nu_1, \nu_2, \nu_3, \ldots \} .
\ee
The elementary prospects are assumed to be mutually disjoint.

\vskip 3mm
{\bf Definition 9. Basic states}

\vskip 2mm
Basic states are the vectors 
\be
\label{7}
|e_n > \;  \equiv \; \otimes_i |A_{i\nu_i} > \; \equiv \; 
|A_{i\nu_1} A_{i\nu_2} \ldots > \; ,
\ee
which are mapped to the elementary prospects labelled in (\ref{6}). 
These vectors are the tensor products of the mode states $|A_{i\nu_i}>$.

\vskip 3mm
{\bf Definition 10. Prospect basis}

\vskip 2mm
The prospect basis $\{ |e_n>\}$ is the family of all basic states (\ref{7})
corresponding to the elementary prospects. Different states belonging to 
the prospect basis are assumed to be disjoint, in the sense of being 
orthogonal. Since the modulus of each state has no special meaning, these 
states are also normalized to one. This can be formalized as the 
orthonormality of the basis, for which there exists a scalar product
\be
\label{8}
< e_m | e_n > \; = \prod_i < A_{i\mu_i} | A_{i\nu_i} > \; = \; \dlt_{mn} \; ,
\ee
where
\be
\label{9}
\dlt_{mn} \equiv \prod_i \dlt_{\mu_i \nu_i}
\ee
is the product of the Kronecker symbols.

\vskip 3mm
{\bf Definition 11. Mind space}

\vskip 2mm
The space of mind is defined as the closed linear envelope over the
prospect basis $\{|e_n>\}$:
\be
\label{10}
\cM \equiv \overline \cL \{ |e_n> \} = \otimes_i \cM_i~.
\ee
This is a Hilbert space, being the direct product of the mode spaces
(\ref{2}), which can be thought of as a possible mathematical 
representation of the mind. Note that the closed linear envelope 
(\ref{10}) exhausts all possible states that can be 
expanded over the total basis $\{|e_n>\}$. Mathematically, 
$\overline\cL\{|e_n>\}$ is identical to $\otimes_i\cM_i$. Therefore 
the product $\otimes_i\cM_i$ is a direct consequence of the structure 
of $\overline\cL\{|e_n>\}$.

\vskip 3mm
{\bf Definition 12. Mind dimensionality}

\vskip 2mm
The dimensionality of the mind space (\ref{10}), which can be
termed the {\it dimensionality of mind}, is
\be
\label{11}
dim(\cM) \; \equiv \; \prod_i M_i \; ,
\ee
where $M_i$ is the number of the $i$-intention modes.

\vskip 3mm
{\bf Definition 13. Prospect states}

\vskip 2mm
A prospect state $|\pi_j>$ is a member of the mind space (\ref{10}).
The prospects are enumerated with the index $j=1,2,\ldots$. The total
set $\{|\pi_j>\}$ of all prospect states $|\pi_j>$, corresponding to
all admissible prospects, forms a subset of the space of mind. The set
$\{|\pi_j>\}\subset\cM$ can be called the prospect-state set. Note that
the vectors $|\pi_j>$ are not necessarily orthogonal with each other 
and, generally, are not normalized. The normalization condition will be 
formulated for the prospect probabilities to be defined below.

\vskip 3mm
{\bf Definition 14. Strategic state}

\vskip 2mm
The strategic state of mind at time $t$ is a given specific vector
\be
\label{12}
|\psi_s(t)> = \sum_n c_n(t) |e_n> \; ,
\ee
which can be represented as a linear combination of the prospect basic 
states $\{|e_n>\}$. The coefficients $c_n(t)$ are given complex-valued 
functions of time, whose temporal evolution is associated with the 
particular individual and context. The strategic state (\ref{12}) belongs 
to the mind space (\ref{10}), which is a Hilbert state endowed with 
the scalar product
\be
\label{13}
< \psi_s(t_1) | \psi_s(t_2) > \; \equiv \sum_n c_n^*(t_1) c_n(t_2) \; .
\ee
The norm of the strategic state (\ref{12}) is generated by the scalar
product (\ref{13}),
\be
\label{14}
|| |\psi_s(t)> || \equiv \sqrt{ <\psi_s(t) |\psi_s(t)> } \; .
\ee
The strategic state of mind is normalized to unity, so that
\be
\label{15}
|| |\psi_s(t)> || = 1 \; .
\ee
Then, from the definition of the scalar product (\ref{13}), we have
\be
\label{16}
\sum_n | c_n(t)|^2 = 1 \; .
\ee
The strategic state of mind is a fixed vector characterizing a particular 
decision maker, with his/her beliefs, habits, principles, etc., that is,
describing each decision maker as a unique subject. Hence, each space of 
mind possesses a unique strategic state. Different decision makers possess 
different strategic states.

\subsection{Entangled Prospects}

Prospect states can be of two qualitatively different types, disentangled 
and entangled.

\vskip 3mm
{\bf Definition 15. Disentangled states}

\vskip 2mm
A disentangled prospect state is a prospect state which is represented as 
the tensor product of the intention states (\ref{3}):
\be
\label{17}
|f(t)> = \otimes_i |\psi_i(t)> \; ,
\ee
We define the {\it disentangled set} as the collection of all admissible 
disentangled prospect states of form (\ref{17}):
\be
\label{18}
\cD \equiv \{ |f> = \otimes_i |\psi_i>, \; |\psi_i> \in \cM_i\}~.
\ee

\vskip 3mm
{\bf Definition 16. Entangled states}

\vskip 2mm
An entangled prospect state is any prospect state that cannot be reduced to 
the tensor product form of disentangled prospect states (\ref{17}).

\vskip 3mm

In quantum theory, it is possible to construct various entangled and
disentangled states (see, e.g., Yukalov, 2003). For the purpose of 
developing a theory of decision making, let us illustrate the above 
definitions by an example of a prospect consisting of two intentions 
with two representations each. Let us consider the prospect of the 
following two intentions: ``to get married" and ``to become rich". And 
let us assume that the intention ``to get married" consists of two 
representations, ``to marry $A$", with the representation state $|A>$, 
and ``to marry $B$", with the representation state $|B>$. And let the 
intention ``to become rich" be formed by two representations, ``to become 
rich by working hard", with the representation state $|W>$, and ``to 
become rich by being a gangster", with the representation state $|G>$. 
Thus, there are two intention states of type (\ref{3}),
\be
\label{19}
|\psi_1> = a_1|A> + a_2|B>  \; , \qquad |\psi_2> = b_1|W> + b_2|G> \; .
\ee
The general prospect state has the form
\be
\label{20}
|\pi > = c_{11}|AW> + c_{12}|AG> + c_{21}|BW> + c_{22}|BG> \; ,
\ee
where the coefficients $c_{ij}$ belong to the field of complex numbers.

Depending on the values of the coefficients $c_{ij}$, the prospect 
state (\ref{20}) can be either disentangled or entangled. If it is 
disentangled, it must be of the tensor product type (\ref{17}), which 
for the present case reads
\be
\label{21}
|f> = |\psi_1> \otimes |\psi_2> = a_1 b_1|AW> + a_1 b_2|AG> +
a_2b_1|BW> + a_2b_2|BG> \; .
\ee
Both states (\ref{20}) and (\ref{21}) include four elementary-prospect 
states (\ref{7}):
\begin{itemize}
\item 
``to marry $A$ and to work hard", $|AW>$,
\item 
``to marry $A$ and become a gangster", $|AG>$,
\item 
``to marry  $B$ and to work hard", $|BW>$,
\item 
``to marry $B$ and become a gangster",  $|BG>$.
\end{itemize}

However, the structure of states (\ref{20}) and (\ref{21}) is different. 
The prospect state (\ref{20}) is more general and can be reduced to 
state (\ref{21}), but the opposite may not be possible. For instance, 
the prospect state
\be
\label{22}
c_{12}|AG> + c_{21}|BW> \; ,
\ee
which is a particular example of state (\ref{20}) cannot be reduced to
any of the states (\ref{21}), provided that both coefficients $c_{12}$ and
$c_{21}$ are non-zero. In quantum mechanics, this state would be called the
Einstein-Podolsky-Rosen state, one of the most famous examples of an
entangled state (Einstein et al., 1935). Another example is the prospect
state
\be
\label{23}
c_{11}|AW > + c_{22}| BG > \; ,
\ee
whose quantum-mechanical analog would be called the Bell state (Bell, 1964).
In the case where both $c_{11}$ and $c_{22}$ are non-zero, the Bell state
cannot be reduced to any of the states (\ref{21}) and is thus entangled.

In contrast with the above two examples, the prospect states
$$
c_{11}|AW> + c_{12}|AG> \; ,  \qquad c_{11}|AW> + c_{21}|BW> \; ,
$$
$$
c_{12}|AG> + c_{22}|BG> \; ,  \qquad c_{21}|BW> + c_{22}|BG> \; ,
$$
are {\it disentangled}, since all of them can be reduced to the
form (\ref{21}).

Since the coefficients $c_{ij}=c_{ij}(t)$ are, in general, functions
of time, it may happen that a prospect state at a particular time is
entangled, but becomes disentangled at another time or, vice versa, a
disentangled prospect state can be transformed into an entangled state
with changing time (Yukalov, 2003).

The state of a human being is governed by his/her physiological 
characteristics and the available information (Bechara et al., 2000; 
Dickhaut et al., 2003). These properties are continuously changing in 
time. Hence the strategic state (\ref{12}), specific of a person 
at a given time, may also display temporal evolution, according to 
different homeostatic processes adjusting the individual to the 
changing environment.

\subsection{Decision Making}

We describe the process of decision making as an intrinsically
probabilistic procedure. The first step consists in evaluating consciously 
and/or subconsciously the probabilities of choosing different actions from 
the point of view of their usefulness and/or appeal to the choosing agent. 
Mathematically, this is described as follows.

\vskip 3mm
{\bf Definition 17. Prospect set}

\vskip 2mm
The total family 
\be
\label{24}
{\cal L} \equiv  \{\pi_j: \; j = 1,2, \ldots \} 
\ee
of all prospects $\pi_j$, among which one makes a choice, is called the 
prospect set.

\vskip 3mm
{\bf Definition 18. Prospect operators}

\vskip 2mm
The prospect operator, corresponding to a prospect $\pi_j$ with the 
prospect state $|\pi_j>$ is
\be
\label{25}
\hat{P}(\pi_j) \equiv |\pi_j> <\pi_j| \;  .
\ee

\vskip 3mm
The prospect operators in decision theory are analogous to the operators 
of local observables in quantum theory. The prospect probabilities are 
defined as the expectation values of the prospect operators with respect 
to the given strategic state. The strategic state of mind of an agent at 
some time $t$ is represented by the state $|\psi_s(t)>$.

\vskip 3mm
{\bf Definition 19. Prospect probabilities}

\vskip 2mm
The probability of realizing a prospect $\pi_j$, with the prospect state 
$|\pi_j>$, under the given strategic state $|\psi_s(t)>$, characterizing 
the agent's state of mind at time $t$, is the expectation value of the 
prospect operator (\ref{25}):
\be
\label{26}
p(\pi_j) \equiv <\psi_s(t)|\hat{P}(\pi_j)|\psi_s(t)> = 
| < \pi_j|\psi_s(t)> |^2 \; .
\ee

The prospect probabilities defined in (\ref{26}) are assumed to possess 
all standard probability properties, with the normalization condition
\be
\label{27}
\sum_j p(\pi_j) = 1 \;  \qquad (0 \leq p(\pi_j) \leq 1 ) \;  .
\ee

\vskip 3mm
The prospect probabilities are defined in Eq. (\ref{26}) through the 
prospect states and the strategic state of mind. The latter is 
normalized to one, according to Eq.(\ref{15}). By their definition, 
the prospect probabilities have to be summed to one, as in Eq. (\ref{27}). 
But the prospect states themselves do not need to be normalized to one, 
which means that different prospects can have, and usually do have, 
different weights, corresponding to their different probabilities. In 
physics, this situation would be similar to defining the cross-section 
in a scattering experiment over a system containing elementary particles 
(elementary prospects) and composite clusters (composite prospects) 
formed by several particles.

In the traditional theory of decision making, based on the utility 
function, the optimal decision corresponds, by definition, to the 
maximal expected utility which is associated with the maximal 
anticipated usefulness and profit resulting from the chosen action.
In contrast, our QDT recognizes that the behavior of an individual
is probabilistic, not deterministic. The prospect probability 
(\ref{26}) quantifies the probability that a given individual chooses 
the prospect $\pi_j$, given his/her strategic state of 
mind $|\psi_s(t)>$ at time $t$. This translates in experiments into a 
prediction on the frequency of the decisions taken by an ensemble of 
subjects under the same conditions. The observed frequencies of 
different decisions taken by an ensemble of non-interacting subjects 
making a decision under the same conditions serves as the observable 
measure of the subjective probability. It is, actually, the known fact
that subjective probabilities can be calibrated by frequencies or 
fractions (Tversky and Kahneman, 1973; Kaplan and Garrick, 1981). 

This specification also implies that the same subject, prepared under 
the same conditions with the same strategic state of mind $|\psi_s>$ at 
two different times, may choose two different prospects among the same 
set of prospects, with different relative frequencies determined by the 
corresponding prospect probabilities (\ref{26}). Verifying this prediction 
is a delicate empirical question, because of the possible impact of the 
``memory" of the past decisions on the next one. In order for the 
prediction to hold, the two repetitions of the decision process should be 
independent. Otherwise, the strategic state of mind in the second 
experiment keeps a memory of the previous choice, which biases the results. 
This should not be confused with the fact that the projection of the 
strategic state of mind onto the prospect state $|\pi_j>$, when the 
decision is made to realize this prospect, ensures that the individual 
will in general keep his/her decision, whatever it is, when probed a 
second time sufficiently shortly after the first decision so that the 
strategic state of mind, realized just after the projection, has not had 
time yet to evolve appreciably.

\vskip 3mm
{\bf Definition 20. Optimal prospect}

\vskip 2mm
The prospect $\pi_*$ is called optimal if and only if its probability is
the largest among the probabilities of all prospects from the considered 
prospect set $\cal{L}$,
\be
\label{28}
p(\pi_*) = \max_j p(\pi_j) \;  \qquad  (\pi_j \in \cal{L})  \;  .
\ee

\vskip 3mm
In QDT, the concept of an optimal decision is replaced by a probabilistic 
decision, when the prospect, which makes the probability $p(\pi_j)$ given 
by (\ref{26}) maximal, is the one which corresponds best to the given 
strategic state of mind of the decision maker. In that sense, the 
prospect which makes $p(\pi_j)$ maximal can be called ``optimal with 
respect to the strategic state of mind". Using the mapping between the 
subjective probabilities and the frequentist probabilities observed on 
ensembles of individuals, the prospect that makes $p(\pi_j)$ maximal will 
be chosen by more individuals than any other prospect, in the limit of 
large population sampling sizes. However, other less probable prospects 
will also be chosen by some smaller subsets of the population.

\vskip 3mm
{\bf Remark 1. Entangled decision making}

\vskip 2mm
As is explained above, a prospect state $|\pi_j>$ does not have in general 
the form of the product (\ref{17}), which means that it is entangled. The 
strategic state $|\psi_s>$ can also be entangled. Therefore, the prospect 
probability $p(\pi_j)$, in general, cannot be reduced to a product of terms,
but has a more complicated structure, as will be shown below. In other 
words, the decision making process is naturally entangled.

Consider the example of Section 2 of the specific prospect state
(\ref{20}) associated with the two intentions ``to get married''
and ``to become rich". And suppose that $A$ does not like gangsters, so
that it is impossible to marry $A$ and at the same time being a gangster.
This implies that the prospect-representation $AG$ cannot be realized,
hence $c_{12}=0$. Assume that $B$ dreams of becoming rich as fast as
possible, and a gangster spouse is much more luring for $B$ than a dull
person working hard, which implies that $c_{21}=0$. In this situation, the
prospect state (\ref{20}) reduces to the entangled Bell state 
$c_{11}|AW>+c_{22}|BG>$. A decision performed under these conditions,
resulting in an entangled state, is entangled.

\vskip 3mm
{\bf Remark 2. Noncommutativity of subsequent decisions}
 
\vskip 2mm
There exist numerous real-life examples when decision makers fail to follow
their plans and change their mind simply because they experience different
outcomes on which their intended plans were based. This change of plans
after experiencing particular outcomes is the effect known as dynamic
inconsistency (Frederick et al., 2002; Barkan et al., 2005; Yukalov and 
Sornette, 2009). In our language, this is a simple consequence of the 
non-commutativity of subsequent decisions, resulting from entanglement 
between intention representations and caused by the existence of intention 
interference.

\section{Prospect Interference}

As soon as one accepts the description of decision making, which invokes the 
mathematical techniques of quantum theory as is suggested by Bohr (1929, 
1933, 1937, 1961), one inevitably meets the effects of interference. The 
possible occurrence of interference in the problems of decision making 
has been mentioned before on formal grounds (see, e.g., Busemeyer et al., 
2006). However, no general theory has been suggested, which would explain 
why and when such effects would appear, how to predict them, 
and how to give a quantitative analysis of them. In our approach, 
interference in decision making arises only when one takes a decision 
involving composite prospects. The corresponding mathematical treatment 
of these interferences within QDT is presented in the following subsections.

\subsection{Illustration of Interference in Decision Making}

As an illustration, let us consider the following situation of two
intentions, ``to get a friend" and ``to become rich". Let the former
intention have two representations ``to get the friend $A$" and ``to 
get the friend $B$''. And let the second intention also have two 
representations, ``to become rich by working hard" and ``to become 
rich by being a gangster''. The corresponding strategic mind state is 
given by Eq. (\ref{12}), with the evident notation for the basic states 
$|e_n>$ and the coefficients $c_{ij}$ given by the identities
$$
c_{11} \equiv c_A(W) \; , \qquad c_{12} \equiv c_A(G) \; , \qquad
c_{21} \equiv c_B(W) \; , \qquad c_{22} \equiv c_B(G) \; .
$$

Suppose that one does not wish to choose between these two friends 
in an exclusive manner, but one hesitates of being a friend to $A$ 
as well as $B$, with the appropriate weights. This means that one 
deliberates between the intention representations $A$ and $B$, while 
the way of life, either to work hard or to become a gangster, has not 
yet been decided.

The corresponding composite prospects
\be
\label{29}
\pi_A = A (W + G) \; , \qquad \pi_B = B (W + G) 
\ee
are characterized by the prospect states
\be
\label{30}
|\pi_A > \; = \; a_1|AW > + a_2|AG > \; , \qquad
|\pi_B > \; = \; b_1|BW > + b_2|BG > \; .
\ee
The coefficients of the prospect states define the weights corresponding 
to the intended actions, among which the choice is yet to be made. One 
should not confuse the intended actions with the actions that have already 
been realized. One can perfectly deliberate between keeping this or that 
friend, in the same way, as one would think about marrying $A$ or $B$ in 
another example above. This means that the choice has not yet been made. 
And before it is made, there exist deliberations involving stronger or 
weaker intentions to both possibilities. Of course, one cannot marry both 
(at least in most Christian communities). But before marriage, there can 
exist the dilemma between choosing this or that individual.

Calculating the scalar products
$$
<\pi_A|\psi_s > \; = \; a_1^* c_{11} + a_2^* c_{12} \; , 
\qquad
<\pi_B|\psi_s > \; = \; b_1^* c_{21} + b_2^* c_{22} \; ,
$$
we find the prospect probabilities
\be
\label{31}
p(\pi_A) = \left | a_1^* c_{11} + a_2^* c_{12} \right |^2 \; , 
\qquad
p(\pi_B) = \left | b_1^* c_{21} + b_2^* c_{22} \right |^2 \; .
\ee

Recall that the prospects are characterized by vectors pertaining to 
the space of mind $\cM$, which are not necessarily normalized to one or 
orthogonal to each other. The main constraint is that the total set of 
prospect states $ {\cal L} = \{|\pi_j>\}$ be such that the related 
probabilities 
$$
p(\pi_j)\equiv|<\pi_j| \psi_s > |^2
$$ 
be normalized to one, according to the normalization condition (\ref{27}).

The probabilities (\ref{31}) can be rewritten in another form by 
introducing the partial probabilities
$$
p(AW) \equiv | a_1 c_{11}|^2 \; , \qquad 
p(AG) \equiv | a_2 c_{12}|^2 \; ,
$$
\be
\label{32}   
p(BW) \equiv | b_1 c_{21}|^2 \; , \qquad 
p(BG) \equiv | b_2 c_{22}|^2 \; ,
\ee
and the interference terms
\be
\label{33} 
q(\pi_A) \equiv 2{\rm Re} \left ( a_1^* c_{11} a_2 c_{12}^* 
\right ) \; , \qquad
q(\pi_B) \equiv 2{\rm Re} \left ( b_1^* c_{21} b_2 c_{22}^* 
\right ) \; .
\ee
Then the probabilities (\ref{31}) become
\be
\label{34} 
p(\pi_A) = p(AW) + p(AG) + q(\pi_A) \; , \qquad
p(\pi_B) = p(BW) + p(BG) + q(\pi_B)  \; .
\ee

Let us define the {\it uncertainty angles} 
\be
\label{35}
\Dlt(\pi_A) \equiv {\rm arg} \left ( a_1^* c_{11} a_2 c_{12}^* 
\right ) \; ,\qquad 
\Dlt(\pi_B) \equiv {\rm arg} \left ( b_1^* c_{21} b_2 c_{22}^* 
\right )
\ee
and the {\it uncertainty factors}
\be
\label{36} 
\vp(\pi_A) \equiv \cos \Dlt(\pi_A) \; , \qquad
\vp(\pi_B) \equiv \cos \Dlt(\pi_B)\;  .
\ee
Using these, the interference terms (\ref{33}) take the form
\be
\label{37}
q(\pi_A) = 2\vp(\pi_A) \; \sqrt{p(AW) p(AG) } \; , \qquad
q(\pi_B) = 2\vp(\pi_B) \; \sqrt{p(BW) p(BG) } \;  .
\ee
The interference terms characterize the existence of deliberations between
the decisions of choosing a friend and, at the same time, a type of work.

This example illustrates the observation that the phenomenon of decision
interference appears when one considers a composite entangled prospect
with several intention representations assumed to be realized simultaneously.
Thus, we can state that interference in decision making appears when one 
decides about a composite entangled prospect.

For the above example of decision making in the case of two
intentions, ``to get a friend" and ``to be rich", the appearance of the
interference can be understood as follows. In real life, it is too
problematic, and practically impossible, to become a very close friend to
several persons simultaneously, since conflict of interests often arises
between the friends. For instance, doing a friendly action to one friend
may upset or even harm another friend. Any decision making, involving
mutual correlations between two persons, necessarily requires taking into
account their sometimes conflicting interests. This is, actually, one of
the origins of the interference in decision making. Another powerful
origin of intention interference is the existence of emotions, as will be
discussed in the following sections.

\subsection{Conditions for the Presence of Interference}

The situations for which intention interferences cannot appear can be
classified into two cases, which are examined below. From this
classification, we conclude that the necessary conditions for the
appearance of intention interferences are that the dimensionality of mind
should be not lower than two and that there should be some uncertainty in 
the considered prospect. These conditions imply that the considered 
prospect can be entangled.

\vskip 3mm
{\bf Case 1. One-dimensional mind}

\vskip 2mm
Suppose there are many intentions $\{ A_i\}$, enumerated by the index
$i=1,2,\ldots$, whose number can be arbitrary. But each intention
possesses only a single representation $|A_i>$. Hence, the dimension 
of ``mind'' is $dim(\cM)=1$. Only a single basic vector exists:
$$
| A_1 A_2 \ldots > \; = \; \otimes_i \; |A_i> \; .
$$
In this one-dimensional mind, all prospect states are disentangled, being
of the type
$$
|\psi> = c\; | A_1 A_2 \ldots > \qquad (|c|=1) \; .
$$
Therefore, only one probability exists:
$$
p = | < A_1 A_2 \ldots | \psi> |^2 = 1 \; .
$$

Thus, despite the possible large number of arbitrary intentions,
they do not interfere, since each of them has just one representation.
There can be no intention interference in one-dimensional mind. 

\vskip 3mm
{\bf Case 2. Absence of uncertainty}

\vskip 2mm
Another important condition for the appearance of intention
interference is the existence of uncertainty. To understand this 
statement, let us consider a given mind with a large dimensionality 
$dim(\cM) > 1$, characterized by a strategic state $|\psi_s>$. Let us 
analyze a certain prospect with the state
$$
|\pi_j> = c_j |\psi_s> \qquad (|c_j|=1) \; .
$$
Then, the corresponding prospect probability is
$$
p(\pi_j) = |<\pi_j|\psi_s>|^2 = 1 \;  ,
$$
and no interference can arise.

Thus, the necessary conditions for the intention interference are the
existence of uncertainty and the dimensionality of mind not lower
than $2$.

\subsection{Interference Alternation}

Let us consider two intentions, one composing a set $\{ A_i\}$ of $M_1$
representations and another one forming a set $\{ X_j\}$ of $M_2$
representations. The total family of intention representations
is therefore
\be
\label{38}
\{ A_i,X_j|\; i=1,2,\ldots,M_1; \; j=1,2,\ldots,M_2\} \; .
\ee
The prospect basis is the set $\{|A_iX_j>\}$. The strategic state of 
mind can be written as an expansion over this basis,
\be
\label{exp39}
|\psi_s> = \sum_{ij} c_{ij} |A_iX_j> \; ,
\ee
with the coefficients satisfying the normalization
\be
\label{40}
\sum_{ij} |c_{ij}|^2 = 1 \; .
\ee

Let us assume that we are mainly interested in the representation set
$\{ A_i\}$, while the representations from the set $\{ X_j\}$ are treated
as additional. A prospect $\pi_i \equiv A_i X$, where $X = \bigcup_i X_i$,
which is formed of a fixed intention representation $A_i$, and which can 
be realized under the occurrence of any of the representations $X_i$, 
corresponds to the prospect state
\be
\label{41}
|\pi_i> = \sum_j a_{ij}|A_i X_j> \; .
\ee
The probability of realizing the considered prospect $\pi_i$ is
\be
\label{42}
p(\pi_i) \equiv | <\pi_i|\psi_s>|^2 \; ,
\ee
according to definition (\ref{26}).

Following the above formalism, used for describing intention 
interferences,  we use the notation
\be
\label{43}
p(A_i X_j) \equiv |a_{ij} c_{ij}|^2
\ee
for the joint probability of $A_i$ and $X_j$; and we denote the
interference terms as
\be
\label{44}
q_{jk}(\pi_i) \equiv 2{\rm Re}\left (
a_{ij}^* c_{ij} c_{ik}^* a_{ik} \right ) \; .
\ee
Then, the probability of $\pi_i$, given by Eq. (\ref{42}), becomes
\be
\label{45}
p(\pi_i) = \sum_j p(A_i X_j) + \sum_{j<k} q_{jk}(\pi_i) \; .
\ee

The interference terms appear due to the existence of uncertainty.
Therefore, we may define the {\it uncertainty angles}
\be
\label{46}
\Dlt_{jk}(\pi_i)
\ee
and the {\it uncertainty factors}
\be
\label{47}
\vp_{jk}(\pi_i) \equiv \cos\Dlt_{jk}(\pi_i) \; .
\ee
Then, the interference terms (\ref{44}) take the form
\be
\label{48}
q_{jk}(\pi_i) = 2\vp_{jk}(\pi_i) \;
\sqrt{p(A_iX_j) \; p(A_iX_k)} \; .
\ee
It is convenient to define the sum of the interference terms
\be
\label{49}
q(\pi_i) \equiv \sum_{j<k} q_{jk}(\pi_i) \; .
\ee
This allows us to rewrite the prospect probability (\ref{45}) as
\be
\label{50}
p(\pi_i) = \sum_j p(A_iX_j) + q(\pi_i) \; .
\ee
The joint and conditional probabilities are related in the standard
way
\be
\label{51}
p(A_iX_j) = p(A_i|X_j) p(X_j) \; .
\ee

In view of the normalization condition (\ref{27}), we have 
$\sum_i p(\pi_i) = 1$, which means that the family of intended actions 
(\ref{38}) is such that at least one of the representations from the 
set $\{ A_i\}$ has to be certainly realized. We also assume that at least 
one of the representations from the set $\{ X_j\}$ necessarily happens, 
that is,
\be
\label{52}
\sum_j p(X_j) = 1 \; .
\ee
Along with these conditions, we keep in mind that at least one of the
representations from the set $\{ A_i\}$ must be realized for each given
$X_j$, which implies that
\be
\label{53}
\sum_i p(A_i|X_j) = 1 \; .
\ee
Then we see that $\sum_i q(A_iX) = 0$.

By introducing the prospect  {\it utility factor}
\be
\label{54}
f(\pi_i) \equiv \sum_j p(A_i X_j) \;    ,
\ee
conditions (\ref{52}) and (\ref{53})
can be combined in one normalization condition
\be
\label{55}
\sum_j f(\pi_j) = 1 \;  .
\ee

The above consideration can be generalized into the following statement.

\vskip 2mm

{\bf Theorem 1. Interference alternation}:
{\it The process of decision making, associated with the 
probabilities $p(\pi_j)$ of the prospects $\pi_j \in \cal{L}$, occurring 
under the normalization conditions (\ref{27}) and (\ref{55}), is 
characterized by alternating interference terms, such that the total 
interference vanishes, which implies the {\it property of interference 
alternation}
\be
\label{56}
\sum_j q(\pi_j) = 0 \; .
\ee
}

{\it Proof}: From the above definitions, it follows that the prospect 
probability has the form
\be
\label{57}
p(\pi_j) = f(\pi_j) + q(\pi_j)  \;  .
\ee 
From here, taking into account the normalization conditions (\ref{27})
and (\ref{55}), we get the alternation property (\ref{56}). 

\vskip 2mm

Equality (\ref{56}) shows that, if at least one of the terms is
non-zero, some of the interference terms are necessarily negative and
some are necessarily positive. Therefore, some of the probabilities are
depressed, while others are enhanced. This alternation of the interference
terms will be shown below to be a pivotal feature providing a clear
explanation of the disjunction effect. It is worth emphasizing that the
violation of the sure-thing principle, resulting in the disjunction effect,
will be shown not to be due simply to the existence of interferences as
such, but more precisely to the {\it interference alternation}.

For instance, the depression of some probabilities can be associated with
uncertainty aversion, which makes less probable an action under uncertain
conditions. In contrast, the probability of other intentions, containing
less or no uncertainty, will be enhanced by positive interference terms.
This interference alternation is of crucial importance for the correct 
description of decision making, without which the known paradoxes cannot 
be explained.

\section{Interference Quarter Law}

In agreement with the form (\ref{57}), the prospect probability $p(\pi_j)$
is the sum of two terms, the utility factor $f(\pi_j)$ and the interference
term $q(\pi_j)$. The first term defines the prospect utility for the decision 
maker. The second term characterizes the prospect attractiveness for
this decision maker, or a subjectively defined prospect quality. Therefore 
the quantity $q(\pi_j)$ can be called the {\it attraction factor} or 
{\it quality factor}. As has been stressed several times throughout the paper, 
this reflects the fact that the interference terms are embodying
subjective feelings and emotions of the decision maker.

The appearance of the interference terms is the consequence of the use of
quantum-theoretical techniques for describing the process of decision 
making. However, the possible occurrence of interference as such does not 
yet provide an explanation of paradoxical effects in human decision 
making. If we would simply postulate the existence of the interference 
terms and would fit them on the basis of some particular experiments,
this would have no scientific value. Our approach may acquire the status
of a theory if (i)  it explains the conditions under which the interference
terms appear, (ii) it delineates their underlying origin and (iii) it provides
a procedure, even approximate, for their quantitative evaluation.
The following proceeds to demonstrate these three points.

\subsection{Aggregate Nature of Quantum Decision Theory}

In the previous sections, we uncovered two important properties of the 
interference terms. First of all, we showed that these terms arise only
when the considered prospects are composite. Second, we derived the
theorem of interference alternation (Theorem 1). These properties clarify
the conditions under which interference can arise. But they are not yet 
sufficient for estimating the values of the interference terms.

Strictly speaking, being defined to reflect subjective 
factors embodying subconscious feelings, emotions, and biases,  the interference terms are
contextual. This means that the values of $q$ can be different for 
different decision makers. Moreover, they can be different for the same 
decision maker at different times. These features seem to be natural 
when one keeps in mind real humans, whose decisions are usually different,
even under identical conditions.
It is also known that the same decision maker can vary his/her decisions
at different times and under different circumstances. But focusing solely on the
contextual character of the interference terms gives the wrong impression 
of a lack of predictive 
power of the approach which would make it rather meaningless.

Fortunately, there is a way around the problem of contextuality, based
on the fact that QDT has been constructed as a 
probabilistic theory, with the probabilities interpreted in the 
frequentist sense. This is equivalent to saying that QDT is a theory
of the aggregate behavior of a population. In other words, the predictions of the theory 
are statistical statements concerning the population of
individualistic behaviors,  namely QDT provides the probability 
for a given individual to take this or that decision interpreted in the sense
of the fraction of individuals taking these decisions. 

Keeping in mind this aggregate nature of QDT, there is no need to discuss
the specific values of the factor $q$ appropriate to particular decision 
makers. But it is necessary to evaluate typical, or expected values of $q$,
corresponding to an ensemble of decision makers under given conditions. In
the following subsections, we show how this can be done. Knowing the 
expected value of $q$ makes it possible to predict the typical behavior 
of decision makers.

\subsection{Binary Prospect Set}

For concreteness, let us consider the case of two prospects. Suppose, one
deliberates between the intended actions $A$ and $B$, under an additional 
intention with two modes, $X = X_1 + X_2$. So that, one chooses between
two composite prospects
\be
\label{58}
\pi_A \equiv AX \;  \qquad   \pi_B = BX  \;   .
\ee
The interference terms (\ref{48}) can be rewritten as
$$
q(\pi_A) = 2\vp(\pi_A) \; \sqrt{p(A|X_1) p(X_1) p(A|X_2) p(X_2)} \; ,
$$
\be
\label{59}
q(\pi_B) = 2\vp(\pi_B) \; \sqrt{p(B|X_1) p(X_1) p(B|X_2) p(X_2)} \; .
\ee
The interference-alternation theorem (Theorem 1), which leads to (\ref{56}), 
implies that
\be
|q(\pi_A)| = |q(\pi_B)| \; ,
\label{60}
\ee
and
\be
{\rm sign}[\vp(\pi_A)] = - {\rm sign}[\vp(\pi_B)]~.
\label{61}
\ee
This defines the relation between the uncertainty factors.

A fundamental well-documented characteristic of human beings is their
aversion to uncertainty, i.e., the preference for known risks over
unknown risks (Epstein, 1999). As a consequence, the propensity/utility
(and therefore the probability) to act under larger uncertainty is smaller
than under smaller uncertainty. Mechanically, this implies that it is 
possible to specify the sign of the uncertainty factors yielding (\ref{61}).

To find the amplitudes of the uncertainty factors, we may proceed as 
follows. By the definition of these factors, we have
\be
\label{62}
| \vp(\pi_A)| \in [0,1] \; , \qquad  | \vp(\pi_B)| \in [0,1]   \; .
\ee 
Without any other information, the simplest prior is to assume a uniform 
distribution of the absolute values of the uncertainty factors in the 
interval $[0,1]$, so that their expected values are respectively
\be
\label{63}
|\overline\vp(\pi_A)| = |\overline\vp(\pi_B)| = \frac{1}{2} \; .
\ee
Choosing in that way the average values of the uncertainty factors is 
equivalent to using a representative agent, while the general approach 
is taking into account a pre-existing heterogeneity. That is, the
values (\ref{63}) should be treated as estimates for the expected 
uncertainty factors, corresponding to these factors averaged with 
the uniform distribution over the large number of agents.

To complete the calculation of $q(\pi_A)$ and of $q(\pi_B)$, given by
(\ref{59}), we also assume the non-informative uniform prior for all
probabilities appearing below the square-roots, so that their expected
values are all $1/2$, since they vary between $0$ and $1$. Using these in
Eq.~(\ref{59}) results in the interference-quarter law
\be
\label{64}
|\overline q(\pi_A)| = |\overline q(\pi_B)| = \frac{1}{4} \; ,
\ee
valid for the four-dimensional mind composed of two intentions with two
representations each.

\subsection{Expected Value of Interference Terms}

In the previous subsection, we have shown that, in the case of a binary 
prospect set, the magnitude of the interference term can be estimated 
by the value $1/4$. Now, we extend this result by demonstrating that 
the expected value of the interference-term magnitude can be estimated
as $1/4$ for an arbitrary prospect, under quite general conditions. 

The interference term, or the attraction factor, $q(\pi_j)$, is defined by 
emotions, subconscious feelings, and other hidden variables. Strictly 
speaking, it is contextual, depending on a particular decision maker 
at a given time. For an ensemble of decision makers, the interference 
term can be treated as a random variable in the interval $[-1,1]$. That 
is, the modulus $|q(\pi_j)|$ of the attraction factor, is a random variable 
in the interval $[0,1]$.

Let the distribution of this random variable be $\rho(\xi)$, with the 
variable $\xi$ in the interval $[0,1]$. The expectation value of the 
modulus of the attraction factor is
\be
\label{65}
q \equiv \int_0^1 \xi \rho(\xi) \; d\xi \; .          
\ee
By its definition, the distribution is normalized as
\be
\label{66}    
\int_0^1 \rho(\xi) \; d\xi = 1 \; .
\ee

Since the exact form of this distribution is not known, we can consider 
two limiting cases. One limiting case is provided by a distribution
concentrated in the center, which is described by the Dirac delta 
function $\delta(\xi)$, so that the distribution is
\be
\label{67}
\rho_1(\xi) = 2 \dlt(\xi) \; .
\ee
Recall that the delta function is defined through the integral
$$
\int_{-a}^a h(\xi) \dlt(\xi) \; d\xi = h(0)  \; ,
$$
where $h(\xi)$ is any smooth function of $\xi$ and $a > 0$. The delta 
distribution is normalized,
$$
 \int_0^1 \rho_1(\xi) \; d\xi = 1 \; . 
$$

Another limiting case is the uniform distribution in the interval 
$[0,1]$, which is described by the form
\be
\label{68}
  \rho_2(\xi) = \Theta(1-\xi) \; ,
\ee 
expressed through the unit-step function
\begin{eqnarray}
\nonumber
\Theta(\xi) \equiv \left \{
\begin{array}{ll}
0, & \xi < 0 \\
1, & \xi > 0. \end{array} \right.
\end{eqnarray}
The uniform distribution is also normalized,
$$
\int_0^1 \rho_2(\xi)\; d\xi = 1 \; .
$$

Knowing only two limiting cases, we may model the unknown 
distribution $\rho(\xi)$ by the average of these two limiting cases:
\be
\label{69}
 \rho(\xi) = \frac{1}{2} \; \left [ \rho_1(\xi) + \rho_2(\xi)
\right ] \; ,
\ee
which yields
\be
\label{70}
 \rho(\xi) = \dlt(\xi) + \frac{1}{2} \; \Theta(1-\xi) \; .
\ee
This distribution, by construction, is normalized as in (\ref{66}). 

Calculating the expected value (\ref{65}), we obtain
\be
\label{71}
  q \equiv \int_0^1 \xi \rho(\xi) \; d\xi = \frac{1}{4} \;  .
\ee
Thus, the expected value of the modulus of the interference term is again 
given by the quarter law: $q = 1/4$. This allows us to quantitatively 
estimate the influence of emotions in decision making and to predict, on
the aggregate level, the average behavior of typical decision makers. 

It is appropriate to remember that it was Bohr (1929, 1933, 1937, 1961) 
who advocated throughout all his life the idea that mental processes do 
bear close analogies with quantum processes. The analogies should be 
understood here in the sense of their similar theoretical description, 
but not necessarily in the sense of being physiologically equivalent. 
Since interference is one of the most striking characteristic features 
of quantum processes, the analogy suggests that it should also arise in 
mental processes as well. The existence of interference in decision making 
disturbs the classical additivity of probabilities. Indeed, we take as an 
evidence of this the nonadditivity of probabilities in psychology which 
has been repeatedly observed (Tversky and Koehler, 1994; Fox et al., 1996; 
Rottenstreich and Tversky, 1997), although it has not been connected with 
interference. 

It is also important to stress that the mere existence of interference 
as such does not allow one to make any reasonable predictions in analyzing 
human decision making. It is necessary to derive the main general 
properties of interference in order to make this notion operationally 
meaningful. These general properties that we have derived are:
\begin{itemize}
\item
Interference appears only for composite prospects under the presence of 
uncertainty.   
\item
Interference terms satisfy the alternation condition formalized in 
Theorem 1.
\item
The expected value of the interference-term magnitude can be estimated 
by the quarter law.
\end{itemize}

Equipped with the knowledge of these properties, it becomes possible 
to analyze the influence of interference on human decision making and explain
the corresponding paradoxical effects.

\section{Disjunction Effect}

The disjunction effect was first specified by Savage (1954) as
a violation of the ``sure-thing principle'', which can be formulated as
follows (Savage, 1954): {\it if the alternative $A$ is preferred to the
alternative $B$, when an event $X_1$ occurs, and it is also preferred
to $B$, when an event $X_2$ occurs, then $A$ should be preferred to
$B$, when it is not known which of the events, either $X_1$ or $X_2$,
has occurred}.

\subsection{Sure-Thing Principle}

For the purpose of self-consistency, let us recall the relationship between
the sure-thing principle and classical probability theory. Let 
us consider a field of events $\{ A,B,X_j|j=1,2,\ldots\}$ equipped 
with the classical probability measures (Feller, 1970). We denote 
the classical probability of an event $A$ by the capital letter $P(A)$ 
in order to distinguish it from the probability $p(A)$ defined in the 
previous sections by means of quantum rules. We shall denote, as usual, 
the conditional probability of $A$ under the knowledge of $X_j$ by 
$P(A|X_j)$ and the joint probability of $A$ and $X_j$, by $P(AX_j)$. 
We assume that at least one of the events $X_j$ from the set 
$\{ X_j\}$ certainly happens, which implies that
\be
\label{72}
\sum_j P(X_j) = 1 \; .
\ee
The probability of $A$, when $X_j$ is not specified, that is, when at 
least one of $X_j$ happens, is denoted by $P(AX)$, with $X=\bigcup_j X_j$.  
The same notations are applied to $B$. Following the common reasoning, 
we understand the statement ``$A$ is preferred to $B$" as meaning 
$P(A)>P(B)$. Then the following theorem is valid.

\vskip 3mm

{\bf Theorem 2}: {\it If for all $j=1,2,\ldots$ one has
\be
\label{73}
P(A|X_j) > P(B|X_j) \; ,
\ee
then
\be
\label{74}
P(AX) > P(BX) \; .
\ee
}

{\it Proof}. Under condition $X = \bigcup_j X_j$, one has
\be
\label{75}
P(AX) = \sum_j P(AX_j) = \sum_j P(A|X_j)P(X_j)
\ee
and
\be
\label{76}
P(BX) = \sum_j P(BX_j) = \sum_j P(B|X_j)P(X_j) \; .
\ee
From Eqs. (\ref{75}) and (\ref{76}), under assumption (\ref{73}),
inequality (\ref{74}) follows immediately.

\vskip 2mm
The above proposition is the theorem of classical probability theory.
Savage (1954) proposed to use it as a normative statement on how human 
beings make consistent decisions under uncertainty. As such, it is no 
more a theorem but a testable assumption about human behavior. In other 
words, empirical tests showing that humans fail to obey the sure-thing 
principle must be interpreted as a failure of humans to abide to the 
rules of classical probability theory.

\subsection{Examples Illustrating the Disjunction Effect}

Thus, according to standard classical probability theory which is held
by most statisticians as the only rigorous mathematical description of
risks, and therefore as the normative guideline describing rational human
decision making, the sure-thing principle should be always verified in 
empirical tests involving real human beings. However, numerous violations 
of this principle have been investigated empirically (Savage, 1954;
Tversky and Shafir, 1992; Croson, 1999; Lambdin and Burdsal, 2007; Li et 
al., 2007). In order to be more specific, let us briefly outline some 
examples of the violation of the sure-thing principle, referred to as 
the disjunction effect.

\vskip 3mm

{\bf Example 1. To gamble or not to gamble}?

\vskip 2mm
A typical setup for illustrating the disjunction effect is a two-step
gamble (Tversky and Shafir, 1992). Suppose that a group of people
accepted a gamble, in which the player can either win an amount of money
(action $X_1$) or lose an amount (action $X_2$). After one gamble, the
participants are invited to gamble the second time, being free to either 
accept the second gamble ($A$) or to refuse it ($B$). Experiments by 
Tversky and Shafir (1992) showed that the majority of people accept the 
second gamble when they know the result of the first one, in any case, 
whether they won or lost in the previous gamble. In the language of 
conditional probability theory, this translates into the fact that people 
act as if $P(A|X_1)$ is larger than $P(B|X_1)$ and $P(A|X_2)$ is larger 
than $P(B|X_2)$ as in Eq. (\ref{73}). At the same time, it turns out that 
the majority refuses to gamble the second time when the outcome of the 
first gamble is not known. The second empirical fact implies that people 
act as if $P(BX)$ overweighs $P(AX)$, in blatant contradiction with 
inequality (\ref{74}) which should hold according to the theorem resulting 
from (\ref{73}). Thus, the majority accepted the second gamble after 
having won or lost in the first gamble, but only a minority accepted the 
second gamble when the outcome of the first gamble was unknown to them. 
This provides an unambiguous violation of the Savage sure-thing principle.

\vskip 3mm

{\bf Example 2. To buy or not to buy}?

\vskip 2mm
Another example, studied by Tversky and Shafir (1992), had to do with
a group of students who reported their preferences about buying a
nonrefundable vacation, following a tough university test. They could
pass the exam ($X_1$) or fail ($X_2$). The students had to decide whether 
they would go on vacation ($A$) or abstain ($B$). It turned out that the 
majority of students purchased the vacation when they passed the exam as 
well as when they had failed, so that condition (\ref{73}) was valid. 
However, only a minority of participants purchased the vacation when 
they did not know the results of the examination. Hence, inequality 
(\ref{74}) was violated, demonstrating again the disjunction effect.

\vskip 3mm

{\bf Example 3. To sell or not to sell}?

\vskip 2mm
The stock market example, analysed by Shafir and Tversky (1992), is a
particularly telling one, involving a deliberation taking into account
a future event, and not a past one as in the two previous cases. Consider
the USA presidential election, when either a Republican wins ($X_1$)
or a Democrat wins ($X_2$). On the eve of the election, market players
can either sell certain stocks from their portfolio ($A$) or hold them
($B$). It is known that a majority of people would be inclined to sell
their stocks, if they would know who wins, regardless of whether the
Republican or Democrat candidate wins the upcoming election. This is
because people expect the market to fall after the elections. Hence,
condition (\ref{73}) is again valid. At the same time, a great many
people do not sell their stocks before knowing who really won the
election, thus contradicting the sure-thing principle and inequality
(\ref{74}). Thus, investors could have sold their stocks before the
election at a higher price, but, obeying the disjunction effect, they
were waiting until after the election, thereby selling at a lower price
after stocks have fallen. Many market analysts believe that this is
precisely what happened after the 1988 presidential election, when George
Bush defeated Michael Dukakis.

\vskip 2mm
There are plenty of other more or less complicated examples of the
disjunction effect (Savage, 1954; Tversky and Shafir, 1992; Shafir
and Tversky, 1992; Shafir et al., 1993; Shafir, 1994;  Croson, 1999;
Lambdin and Burdsal, 2007). The common necessary conditions for the
disjunction effect to arise are as follows. First, there should be
several events, each characterized by several alternatives, as in the
two-step gambles. Second, there should necessarily exist some uncertainty,
whether with respect to the past, as in Examples 1 and 2, or with 
respect to the future, as in Example 3.

Several ways of interpreting the disjunction effect have been analyzed.
Here, we do not discuss the interpretations based on the existence of
some biases, such as the gender bias, or which invoke the notion of
decision complexity, which have already been convincingly ruled out
(Croson, 1999; K\"uhberger et al., 2001). We describe the reason-based
explanation which appears to enjoy a wide-spread following and discuss 
its limits before turning to the view point offered by QDT.

\subsection{Reason-Based Analysis}

The dominant approach for explaining the disjunction effect is the
reason-based analysis of decision making (Tversky and Shafir, 1992;
Shafir and Tversky, 1992; Shafir et al., 1993; Shafir, 1994; Croson,
1999). This approach explains choice in terms of the balance between
reasoning for and against the various alternatives. The basic intuition
is that when outcomes are known, a decision maker may easily come up
with a definitive reason for choosing an option. However, in case of
uncertainty, when the outcomes are not known, people may lack a clear
reason for choosing an option and consequently they abstain and
make an irrational choice.

From our perspective, the weakness of the reason-based analysis is that
the notion of ``reason" is too vague and subjective. Reasons are not only
impossible to quantify, but it is difficult, if possible at all, to give a
qualitative definition of what they are. 

Consider Example 1 ``to gamble
or not to gamble?"  Suppose you have already won at the first step.
Then, you can rationalize that gambling a second time is not very risky:
if you now lose, this loss will be balanced by the first win (on which you
were not counting anyway, so that you may actually treat it differently
from the rest of your wealth, according to the so-called ``mental
accounting'' effect), and if you win again, your profit will be doubled.
Thus, you have a ``reason'' to justify the attractiveness of the second
gamble. But, it seems equally justified to consider the alternative
``reason'':  if you have won once, winning the second time may seem less
probable (the so-called gambler's fallacy), and if you lose, you will
keep nothing of your previous gain. This line of reasoning justifies to
keep what you already got and to forgo the second gamble. Suppose now you
have lost in the first gamble and know it. A first reasoning would be that
the second gamble offers a possibility of getting out of the loss, which
provides a reason for accepting the second gamble. However, you may also
think that the win is not guaranteed, and your situation could actually
worsen, if you lose again. Therefore, this makes it more reasonable not
to risk so much and to refrain from the new gamble. 

Consider now the
situation where you are kept ignorant of whether you have won or lost
in the first gamble. Then, you may think that there is no reason and
therefore no motivation for accepting the second gamble, which is the
standard reason-based explanation. But, one could argue that it would be
even more logical if you would think as follows:  Okay, I do not know what
has happened in the first gamble. So, why should I care about it? Why don't
I try again my luck? Certainly, there is a clear reason for gambling that
could propagate the drive to gamble the second time.

This discussion is not pretending to demonstrate anything other than that 
the reason-based explanation is purely ad-hoc, with no real explanatory 
power; it can be considered in a sense as a reformulation of the disjunction
fallacy. It is possible to multiply the number of examples demonstrating
the existence of quite ``reasonable'' justifications for doing something
as well as a reason for just doing the opposite. It seems to us that the
notion of ``reason" is not well defined and one can always invent in this
way a justification for anything. Thus, we propose that the disjunction
effect has no direct relation to reasoning. In the following section, we
suggest another explanation of this effect based on QDT, specifically the
negative interference between the two uncertain outcomes resulting from
an aversion to uncertainty (uncertainty-aversion principle), which
provides a {\it quantitative} testable prediction.

\subsection{Quantitative Analysis within Quantum Decision Theory}

The possibility of connecting the violation of the sure-thing principle 
with the occurrence of interference has been mentioned in several articles 
(see, e.g., Busemeyer et al. (2006). But these attempts were just ad hoc 
assumptions not based on a self-consistent theory. Our explanation of the 
disjunction effect differs from these attempts in several aspects. First, 
we consider the disjunction effect as just one of several possible effects 
in the frame of the {\it general theory}. The explanation is based on the
theorem of {\it interference alternation}, which has never been mentioned, 
but without which no explanation can be complete and self-consistent. We
stress the importance of the {\it uncertainty-aversion principle}. Also,
we offer a {\it quantitative estimate} for the effect, which is principally 
new.

\subsubsection{Application to Examples of the Disjunction Effect}

Let us discuss the two first examples illustrating the disjunction effect,
in which the prospect consists of two intentions with two representations
each. One intention ``to decide about an action" has the representations
``to act" ($A$) and ``not to act" ($B$). The second intention ``to know
the results" (or ``to have information") has also two representations. One
($X_1$) can be termed  ``to learn about the win" (gamble won, exam passed),
the other ($X_2$) can be called ``to learn about the loss" (gamble lost,
exam failed).  Given the numbers of these representations $M_1=2$ and 
$M_2=2$, the dimension of mind is $dim(\cM)=M_1 M_2=4$.

For the considered cases, the general set of equations for the prospect
probabilities reduces to two equations
$$
p(AX) = p(AX_1) + p(AX_2) + q(AX) \; ,
$$
\be
\label{77}
p(BX) = p(BX_1) + p(BX_2) + q(BX) \; ,
\ee
in which $X = \bigcup_i X_i$ and the interference terms are
$$
q(AX) = 2\vp(AX) \; \sqrt{p(AX_1)\; p(AX_2) } \; ,
$$
\be
\label{78}
p(BX) = 2\vp(BX) \; \sqrt{p(BX_1)\; p(BX_2) } \; .
\ee
Of course, Eqs. (\ref{77}) and (\ref{78}) could be postulated, but then
it would not be clear where they come from. In QDT, these equations appear
naturally. Here $\vp(AX)$ and $\vp(BX)$ are the uncertainty factors defined 
in (\ref{47}). The normalization conditions become 
\be
\label{79}
p(AX) + p(BX) = 1 \; , \qquad p(X_1) + p(X_2) = 1\; ,
\ee
with conditions (\ref{53}) being
\be
\label{80}
p(A|X_1) + p(B|X_1) = 1 \; , \qquad
p(A|X_2) + p(B|X_2) = 1 \; .
\ee
The uncertainty factors can be rewritten as
\be
\label{81}
\vp(AX) = \frac{q(AX)}{2\sqrt{p(AX_1)p(AX_2)} } \; , \qquad
\vp(BX) = \frac{q(BX)}{2\sqrt{p(BX_1)p(BX_2)} } \; ,
\ee
with the interference terms being
\be
\label{82}
q(AX) = p(AX) - p(AX_1) - p(AX_2) \; , \qquad
q(BX) = p(BX) - p(BX_1) - p(BX_2) \; .
\ee

The principal point is the condition of {\it interference alternation}
(Theorem 1), which now reads
\be
\label{83}
q(AX) + q(BX) = 0 \; .
\ee
Without this condition (\ref{83}), the system of equations for the
probabilities would be incomplete, and the disjunction effect could not
be explained in principle.

In the goal of explaining the disjunction effect, it is not sufficient to
merely state that some type of interference is present. It is necessary to
determine (quantitatively) why the probability of acting is suppressed, 
while that of remaining passive is enhanced. Our aim is to evaluate the 
expected size and sign of the interference terms $q(AX)$ (for acting under 
uncertainty) and $q(BX)$ (for remaining inactive under uncertainty). 
Obviously, it is an illusion to search for a universal value that everybody 
would strictly use. Different experiments with different people have indeed 
demonstrated a significant heterogeneity among people, so that, in the 
language of QDT, this means that the values of the interference terms can 
fluctuate from individual to individual. A general statement should here
refer to the behavior of a sufficiently large ensemble of people, allowing
us to map the observed frequentist distribution of decisions to the 
predicted QDT probabilities.

\subsubsection{Alternation Theorem and Interference-Quarter Law}

Now we shall employ the alternation theorem and the quarter law for 
describing the disjunction effect. The interference terms are given in 
(\ref{59}). The interference-alternation theorem (Theorem 1) yields 
Eqs. (\ref{60}) and (\ref{61}). Hence, in the case where 
$p(A|X_j) > p(B|X_j)$, which is characteristic of the examples 
illustrating the disjunction effect, one must have the uncertainty factors 
which exhibit the opposite property, $|\vp(AX)|< |\vp(BX)|$, so as to 
compensate the former inequality to ensure the validity of equality 
(\ref{60}) for the absolute values of the interference terms. The expected 
values of the latter can be evaluated from the Quarter Law as $1/4$.

The next step is to determine the sign of $\vp(AX)$ and, thus, of 
$\vp(BX)$), from (\ref{61}) and their typical amplitudes $|\vp(AX)|$ and 
$|\vp(BX)|$. A fundamental well-documented characteristic of human beings 
is their aversion to uncertainty, i.e., the preference for known risks 
over unknown risks (Epstein, 1999). As a consequence, the 
propensity/utility and, therefore, the probability to act under larger 
uncertainty is smaller than under smaller uncertainty. Mechanically, this 
implies that it is possible to specify the sign of the uncertainty factors, 
yielding
\be
{\rm sign}[\vp(AX)] = - {\rm sign}[\vp(BX)] <0~,
\label{84}
\ee
since $A$ (respectively $B$) refers to acting (respectively to remaining
inactive).

As a consequence of (\ref{84}) and also of their mathematical
definition (\ref{47}), the uncertainty factors vary in the intervals
\be
\label{85}
-1 \leq \vp(AX) \leq 0 \; , \qquad 0 \leq \vp(BX) \leq 1 \; .
\ee
Invoking the interference-quarter law, we find the expected values of 
the interference terms 
\be
\label{86}
\overline q(AX) = -0.25 \; , \qquad
\overline q(BX) = 0.25 \; .
\ee

As a consequence, the probabilities for acting or for remaining inactive 
under uncertainty, given by (\ref{77}), can be evaluated as
$$
p(AX) = p(AX_1) + p(AX_2) - 0.25 \; ,
$$
\be
\label{87}
p(BX) = p(BX_1) + p(BX_2) + 0.25 \; .
\ee
The influence of intention interference in the presence of uncertainty
on the decision making process at the basis of the disjunction effect can
thus be estimated a priori. The sign of the effect is controlled by the
aversion to uncertainty exhibited by people (uncertainty-aversion
principle). The amplitude of the effect can be estimated, as shown above,
from simple priors applied to the mathematical structure of the QDT
formulation.

\subsubsection{Principle of Uncertainty Aversion}

The above calculation implies that the disjunction effect can be 
interpreted as essentially an emotional reaction associated with the 
{\it aversion to uncertainty}. An analogy can make the point: it is widely 
recognized that uncertainty frightens living beings, whether humans or 
animals. It is also well documented that fear paralyzes, as in the cartoon 
of the ``rabbit syndrome,'' when a rabbit stays immobile in front of an 
approaching boa instead of running away. There are many circumstantial 
evidences that uncertainty may frighten people as a boa frightens rabbits. 
Being afraid of uncertainty, a majority of human beings may be hindered to 
act. In the presence of uncertainty, they do not want to act, so that they 
refuse the second gamble, as in Example 1, or forgo the purchase of a 
vacation, as in Example 2, or refrain from selling stocks, as in Example 3. 
Our analysis suggests that it is the aversion to uncertainty that paralyzes
people and causes the disjunction effect. 

It has been reported that, if people, when confronting uncertainty
paralyzing them against acting, are presented with a detailed explanation
of the possible outcomes, they then may change their mind and decide to
act, thus reducing the disjunction effect (Tversky and Shafir, 1992;
Croson, 1999). Thus, encouraging people to think by providing them
additional explanations, it is possible to influence their minds. In such
a case, reasoning plays the role of a kind of therapeutic treatment
decreasing the aversion to uncertainty. This line of reasoning suggests
that it should be possible to decrease the aversion to uncertainty by
other means, perhaps by distracting people or by taking food, drink or 
drug injections. This provides the possibility to test for the dependence 
of the strength of the disjunction effect with respect to various parameters
which may modulate the aversion response of individuals to uncertainty.

We should stress that our explanation departs fundamentally from the
standard reason-based rationalization of the disjunction effect summarized
above. Rather than using what we perceive is an hoc explanation, we anchor 
the disjunction effect on the very fundamental characteristic of living 
beings, that of the aversion to uncertainty. This allows us to construct a 
robust and parsimonious explanation. But this explanation arises only 
within QDT, because QDT allows us to account for the complex emotional, 
often subconscious, feelings as well as the many unknown states of nature 
that underlie decision making. Such unknown states, analogous to hidden 
variables in quantum mechanics, are taken into account by the formalism of 
QDT through the interference alternation effect, capturing mental 
processes by means of quantum-theory techniques.

\subsubsection{Numerical Analysis of Disjunction-Effect Examples}

Let us now turn to the examples described above and suggest their 
quantitative explanations.

\vskip 3mm
{\it Example 1. To gamble or not to gamble}?

\vskip 2mm
The statistics reported by Tversky and Shafir (1992) are
$$
p(A|X_1)=0.69 \; , \qquad p(A|X_2)=0.59 \; , \qquad p(AX)=0.36 \; .
$$
Then Eqs. (\ref{79}) and (\ref{80}) give
$$
p(B|X_1)=0.31 \; , \qquad p(B|X_2)=0.41 \; , \qquad p(BX)=0.64 \; .
$$
Recall that the disjunction effect here is the violation of the sure-thing
principle, so that, although $p(A|X_j)>p(B|X_j)$ for $j=1,2$, one observes
nevertheless that $p(AX)<p(BX)$. In the experiment reported by Tversky and
Shafir (1992), the probabilities for winning or for losing were identical:
$p(X_1)=p(X_2)=0.5$. Then, using relation (\ref{51}), we obtain
$$
p(AX_1) =0.345\; , \qquad p(AX_2)=0.295\; , \qquad
p(BX_1) =0.155\; , \qquad p(BX_2)=0.205 \; .
$$
For the interference terms, we find
\be
q(AX) = -0.28\; , \qquad q(BX) = 0.28 \; .
\label{88}
\ee
The uncertainty factors (\ref{81}) are therefore
$$
\vp(AX) = -0.439 \; , \qquad \vp(BX) = 0.785 \; .
$$
They are of opposite sign, in agreement with condition (\ref{83}). The 
probability $p(AX)$ of gambling under uncertainty is suppressed by the
negative interference term $q(AX)<0$. Reciprocally, the probability
$p(BX)$ of not gambling under uncertainty is enhanced by the positive
interference term $q(BX)>0$. This results in the disjunction effect, 
when $p(AX)<p(BX)$). 

It is important to stress that the observed amplitudes in (\ref{88}) 
are close to the value $0.25$ predicted by the interference-quarter 
law. They are, actually, undistinguishable from $0.25$ within the 
typical statistical error of $20 \%$ characterizing these experiments. That is, even not knowing the 
results of the considered experiment, we are able to {\it quantitatively} 
predict the strength of the disjunction effect.

\vskip 3mm
{\it Example 2. To buy or not to buy}?

\vskip 2mm
For the second example of the disjunction effect, the data, taken from 
Tversky and Shafir (1992), read
$$
p(A|X_1)=0.54 \; , \qquad p(A|X_2)=0.57 \; , \qquad p(AX) = 0.32 \; .
$$
Following the same procedure as above, we get
$$
p(B|X_1)=0.46 \; , \qquad p(B|X_2)=0.43 \; , \qquad p(BX) = 0.68 \; .
$$
Given again that the two alternative outcomes are equiprobable,
$p(X_1)=p(X_2)=0.5$, we find
$$
p(AX_1) =0.270\; , \qquad p(AX_2)=0.285\; , \qquad
p(BX_1) =0.230\; , \qquad p(BX_2)=0.215 \; .
$$
For the interference terms, we obtain
\be
q(AX) = - 0.235 \; , \qquad q(BX) = 0.235 \; .
\label{89}
\ee
The uncertainty factors are
$$
\vp(AX)=-0.424 \; , \qquad \vp(BX)=0.528 \; .
$$
Again, the values obtained in (\ref{89}) are close to those predicted by 
the interference-quarter law, being undistinguishable from $0.25$ within 
experimental accuracy.

Because of the uncertainty aversion, the probability $p(AX)$ of 
purchasing a vacation is suppressed by the negative interference 
term $q(AX)<0$. At the same time, the probability $p(BX)$ of not 
buying a vacation under uncertainty is enhanced by the positive 
interference term $q(BX)>0$. This alternation of interferences 
causes the disjunction effect, when $p(AX)<p(BX)$. It is necessary to 
stress it again that without this interference alternation no 
explanation of the disjunction effect is possible in principle.

In the same way, our approach can be applied to any other situation
related to the disjunction effect associated with the violation of the
sure-thing principle.

\section{Conjunction Fallacy}

The conjunction fallacy constitutes another example revealing that
intuitive estimates of probability by human beings do not conform to the
standard probability calculus. This effect was first studied by Tversky
and Kahneman (1980, 1983) and then discussed in many other works (see,
e.g., Morier and Borgida, 1984; Wells, 1985; Yates and Carlson, 1986;
Shafir et al., 1990; Tentori et al., 2004). Despite an extensive debate
and numerous attempts to interpret this effect, there seems to be no
consensus on the origin of the conjunction fallacy (Tentori et al., 2004).

Here, we show that this effect finds a natural explanation in QDT. It is 
worth emphasizing that we do not invent a special scheme for this 
particular effect, but we show that it is a natural consequence of the 
general theory we have developed. In order to claim to explain the 
conjunction fallacy in terms of an interference effect in a quantum 
description of probabilities, it is necessary to derive the quantitative 
values of the interference terms, amplitudes and signs, as we have done 
above for the examples illustrating the disjunction effect. This has never 
been done before. Our QDT provides the necessary ingredients, in terms of 
the uncertainty-aversion principle, the theorem on interference alternation, 
and the interference-quarter law. Only the establishment of these general 
laws can provide an explanation of the conjunction fallacy, that can be 
taken as a positive step towards validating QDT, according to the general 
methodology of validating theories (Sornette et al., 2007). Finally, in 
our comparison with available experimental data, we analyze a series of 
experiments and demonstrate that all their data substantiate the validity 
of the general laws of the theory.

\subsection{Individual versus Group Decisions}

In order to be precise, it is necessary to distinguish the conjunction 
fallacy observed in the process of decision making performed by separate individuals
and by groups of decision makers. Group decisions can be 
different from those of noninteracting individuals (Baron, 1998; Sheremeta
and Zhang, 2009). In particular, the conjunction fallacy, that has been
documented for isolated decision makers, practically disappears
for decisions taken by groups of interacting 
individuals. The violation rate characterizing the conjunction fallacy falls 
significantly when communication between participants is allowed 
(Charness et al., 2008). The reduction of the strength of the conjunction 
effect is due to the existence of social interactions. These social interactions
play a role analogous to the interaction between particles, which are known
to lead to ``decoherence'' in quantum systems. A study of the
decoherence phenomenon in the present context is beyond the scope
of our paper, which focuses on the conjunction fallacy associated with
separate individuals, in absence of social interactions. This corresponds to
the setup that was studied by Tversky and Kahneman (1980, 1983).

\subsection{Conjunction Rule}

Let us first briefly recall the conjunction rule of standard probability
theory. Let us consider an event $A$ that can occur together with another
one among several other events $X_j$, where $j=1,2,\ldots$. The probability
of an event estimated within classical probability theory is again denoted
with the capital letter $P(A)$, to distinguish it from the probability $p(A)$
in our quantum approach. According to standard probability theory (Feller,
1970), one has
\be
\label{90}
P(AX) = \sum_j P(AX_j) \; ,
\ee
where $X=\bigcup_i X_i$. Since all terms in the sum (\ref{90}) are positive, 
the conjunction rule tells us that
\be
\label{91}
P(AX) \geq P(AX_j) \;  ~~~~(\forall ~ j)~.
\ee
That is, the probability for the occurrence of the conjunction of two
events is never larger than the probability for the occurrence of a 
separate event.

\subsection{Conjunction Error}

Counterintuitively, humans rather systematically violate the conjunction
rule (\ref{91}), commonly making statements such that
\be
\label{92}
p(AX) < p(AX_j) \; ,
\ee
for some $j$, which is termed the {\it conjunction fallacy} (Tversky and 
Kahneman, 1980; 1983). The difference
\be
\label{93}
\ep (AX_j) \equiv p(AX_j) - p(AX)
\ee
is called the {\it conjunction error}, which is positive under conditions
in which the conjunction fallacy is observed.

A typical situation is when people judge about a person, who can possess 
a characteristic $A$ and also some other characteristics $X_j$
(which can be ``possessing a trait'' or ``not having the trait'', since
not having a trait is also a characteristic), as in the oft-cited example
of Tversky and Kahneman (1980): ``Linda is 31 years old, single,
outspoken, and very bright. She majored in philosophy. As a student, she
was deeply concerned with issues of discrimination and social justice,
and also participated in anti-nuclear demonstrations. Which is more likely?
(i) Linda is a bank teller; (ii) Linda is a bank teller and is active in
the feminist movement.''  Most people answer (ii) which is an example of
the conjunction fallacy (\ref{92}). Numerous other examples of the
fallacy are described in the literature (Tversky and Kahneman, 1980, 1983;
Morier and Borgida, 1984; Wells, 1985; Yates and Carlson, 1986; Shafir et
al., 1990; Tentori et al., 2004). It is important to stress that this
fallacy has been reliably and repeatedly documented, that it cannot be
explained by the ambiguity of the word ``likely" used in the formulation
of the question, and that it appears to involve a failure to coordinate
the logical structure of events in the presence of chance (Tentori et al.,
2004).

\subsection{Conjunction Interference}

Within QDT, the conjunction fallacy finds a simple and natural
explanation. Let us consider a typical situation of the fallacy, when
one judges a person who may have a characteristic $A$, treated as primary,
and who may also possess, or not possess, another characteristic, labelled
as secondary. Generally, the person could also be an object, a fact, or
anything else, which could combine several features. Translating this
situation to the mathematical language of QDT, we see that it involves
two intentions. One intention, with just one representation, is ``to decide
whether the object has the feature $A$.'' The second intention ``to decide
about the secondary feature" has two representations, when one decides
whether ``the object has the special characteristic" $(X_1)$ or ``the
object does not have this characteristic" $(X_2)$.

For these definitions, and following the general scheme, we have
\be
\label{94}
p(AX) = p(AX_1) + p(AX_2) + q(AX) =
p(A|X_1)p(X_1) + p(A|X_2)p(X_2) + q(AX) \; 
\ee
where $X=\bigcup_i X_i$. This is a typical situation where a decision is 
taken under uncertainty. The uncertainty-aversion principle imposes that 
the interference term $q(AX)$ should be negative. Taking the perspective 
of the representation $X_1$, definition (\ref{93}) together with 
Eq. (\ref{94}) imply that the conjunction error reads
\be
\label{95}
\ep(AX_1) = | q(AX)| - p(AX_2) \; .
\ee
The condition for the conjunction fallacy to occur is that the error
(\ref{95}) be positive, which requires that the interference term be
sufficiently large, such that the {\it conjunction-fallacy condition}
\be
\label{96}
| q(AX)| > p(AX_2)
\ee
be satisfied.

The QDT thus predicts that a person will make a decision exhibiting
the conjunction fallacy when (i) uncertainty is present and (ii) the
interference term, which is negative by the uncertainty-aversion
principle, has a sufficiently large amplitude, according to 
condition (\ref{96}).

\subsection{Comparison with Experiments}

For a quantitative analysis, we take the data from Shafir et al.
(1990), who present one of the most carefully accomplished and thoroughly
discussed set of experiments. Shafir et al. questioned large groups of
students in the following way. The students were provided with booklets
each containing a brief description of a person. It was stated that the
described person could have a primary characteristic $(A)$, and also
another characteristic $(X_1)$ or its absence $(X_2)$.

In total, there were 28 experiments separated into two groups according
to the conjunctive category of the studied characteristics. In 14 cases,
the features $A$ and $X_1$ were compatible with each other, and in the 
other 14 cases, they were incompatible. The characteristics were treated 
as compatible, when they were felt as closely related according to some
traditional wisdom, for instance, ``woman teacher" $(A)$ and ``feminist"
$(X_1)$. Another example of compatible features is ``chess player" $(A)$ and
``professor" $(X_1)$. Those characteristics that were not related by direct
logical connections were considered as incompatible, such as ``bird watcher"
$(A)$ and ``truck driver" $(X_1)$ or ``bicycle racer" $(A)$ and ``nurse" 
$(X_1)$.

In each of the 28 experiments, the students were asked to evaluate both
the typicality and the probability of $A$ and $AX_1$. Since normal people
usually understand ``typicality'' just as a synonym of probability, and
vice versa, the predictions on typicality were equivalent to estimates of
probabilities. This amounts to considering only how the students estimated
the probability $p(AX)$ that the considered person possesses the stated
primary feature and the probability $p(AX_1)$ that the person has both
characteristics $A$ and $X_1$.

An important quality of the experiments by Shafir et al. (1990) lies
in the large number of tests which were performed. Indeed, a given
particular experiment is prone to exhibit a significant amount of
variability, randomness or ``noise". Not only the interrogated
subjects exhibited significant idiosyncratic differences, with diverse
abilities, logic, and experience, but in addition the questions were
quite heterogeneous. Even the separation of characteristics into two
categories of compatible and incompatible pairs is, to some extent,
arbitrary. Consequently, no one particular case provides a sufficiently
clear-cut conclusion on the existence or absence of the conjunction
fallacy. It is only by realizing a large number of interrogations, with
a variety of different questions, and by then averaging the results,
that it is possible to make justified conclusions on whether or not the
conjunction fallacy exists. The set of experiments performed by Shafir
et al. (1990) well satisfies these requirements.

For the set of compatible pairs of characteristics, it turned out that
the average probabilities were $p(AX)=0.537$ and $p(AX_1)=0.567$, with
statistical errors of $20\%$. Hence, within this accuracy, $p(AX)$ and
$p(AX_1)$ coincide and no conjunction fallacy arises for compatible
characteristics. From the view point of QDT, this is easily interpreted
as due to the lack of uncertainty: since the features $A$ and $X_1$ are
similar to each other, one almost certainly yielding the other, there
is no uncertainty in deciding, hence, no interference, and, consequently,
no conjunction fallacy.

However, for the case of incompatible pairs of characteristics, the
situation was found to be drastically different. To analyse the
related set of experiments, we follow the general scheme of QDT, using the 
same notations as above. We have the prospect with two intentions, one 
intention is to evaluate a primary feature $(A)$ of the object, and another 
intention is to decide whether at the same time the object possesses a 
secondary feature $(X_1)$ or does not possess it $(X_2)$. Taking the data 
for $p(X_j)$ and $p(AX_1)$ from Shafir et al. (1990), we calculate 
$q(AX)$ for each case separately and then average the results. In the 
calculations, we take into account that the considered pairs of 
characteristics are incompatible with each other. The simplest and most 
natural mathematical embodiment of the property of  ``incompatibility'' is 
to take the probabilities of possessing $A$, under the condition of either 
having or not having $X_1$, as equal, that is, $p(A|X_j)=0.5$. For such a 
case of incompatible pairs of characteristics, Eq. (\ref{94}) reduces to
\be
\label{97}
p(AX) = \frac{1}{2} + q(AX) \; .
\ee
The results, documenting the existence of the interference terms
underlying the conjunction fallacy, are presented in Table 1, which
gives the abbreviated names for the object characteristics, whose
detailed description can be found in Shafir et al. (1990).

The average values of the different reported probabilities are
$$
p(AX) = 0.22   \; , \qquad p(X_1)=0.692 \; , \qquad p(X_2)=0.308 \; ,
$$
\be
\label{98}
p(AX_1)=0.346 \; ,  \qquad p(AX_2)=0.154.
\ee
One can observe that the interference terms fluctuate around a mean
of $-0.28$, with a standard deviation of $\pm 0.06$:
\be
\label{99}
\overline q(AX) = - 0.28\pm 0.06 \; .
\ee
There is a clear evidence of the conjunction fallacy, with the
conjunction error (\ref{93}) being $\ep(AX_1)=0.126$.

QDT  interprets the conjunction effect as due to the uncertainty
underlying the decision, which leads to the appearance of the
intention interferences. The interference of intentions is caused 
by the hesitation whether, under the given primary feature $(A)$, 
the object possesses the secondary feature $(X_1)$ or does not have 
it $(X_2)$.  The term $\overline q(AX)$ is negative, reflecting 
the effect of deciding under uncertainty, according to the 
uncertainty-aversion principle. Quantitatively, we observe that 
the amplitude $|\overline q(AX)|$ is in agreement with the QDT  
interference-quarter law, actually coinciding with $0.25$ within 
the experimental accuracy.

\subsection{Conjunction and Disjunction Effects}

The QDT predicts that setups in which the conjunction fallacy occurs
should also be accompanied by the disjunction effect. To see this, let
us extend slightly the previous decision problem by allowing for two
representations of the first intention. Concretely, this means that the
intention, related to the decision about the primary characteristic,
has two representations: (i) ``decide about the object or person
having or not the primary considered feature" $(A)$, and  (ii)  ``decide
to abstain from deciding about this feature" $(B)$. This frames the
problem in the context previously analysed for the disjunction effect. 
The conjunction fallacy occurs when one considers incompatible
characteristics (Tversky and Kahneman, 1983; Shafir et al., 1990),
such that the probabilities of deciding of having a conjunction $(AX_j)$
or of not guessing about it $(BX_j)$ are close to each other, so that one
can set
\be
\label{100}
p(A|X_j) = p(B|X_j) \qquad (\forall j) \; .
\ee
The theorem on interference alternation (Theorem 1) implies that the 
interference term for being passive under uncertainty
is positive and we have
\be
\label{101}
q(BX) = - q(AX) > 0 \; .
\ee
Now, the probability $p(BX)$ of deciding not to guess under uncertainty
is governed by an equation similar to Eq. (\ref{94}). Combining this
equation with (\ref{101}), we obtain
\be
\label{102}
p(BX) = p(AX) + 2 | q(AX) | \; ,
\ee
which shows that, despite equality (\ref{100}), the probability of
being passive is larger than the probability of acting under uncertainty.
This is nothing but a particular case of the disjunction effect.

This example shows that the conjunction fallacy is actually a sufficient
condition for the occurrence of the disjunction effect, both resulting
from the existence of interferences between probabilities under
uncertainty. The reverse does not hold: the disjunction effect does not
necessarily yield the conjunction fallacy, because the latter requires not
only the existence of interferences, but also that their amplitudes would
be sufficiently large according to the conjunction-fallacy condition
(\ref{96}).

To our knowledge, experiments or situations when the disjunction and
conjunction effects are observed simultaneously have not been investigated.
The specific prediction coming from the QDT, that the disjunction effect
should be observable as soon as the conjunction effect is present,
provides a good test of QDT.

We have considered here the case when participants take 
decisions independently, without consulting with each other. When decisions 
are taken in groups, the conjunction fallacy becomes much weaker
(Charness et al., 2008). In the language of QDT, the social interactions
cause the phenomenon of decoherence, which influences the 
strategic state and destroys the interferences.

\section{Conclusion}

In the present paper, we have suggested a quantum theory of decision 
making. By its nature, it can, of course, be realized by a quantum object, 
say, by a quantum computer. Or it can be used as a scheme for quantum 
information processing and for creating artificial intelligence based on 
quantum laws. This, however, is not compulsory. And the developed theory 
can also be applied to non-quantum objects with an equal success. It just 
turns out that the language of quantum theory is a very convenient tool for
describing the process of decision making performed by any decision
maker, whether quantum or not. In this language, it is straightforward
to characterize such features of decision making as the entangled
decision making, non-commutativity of subsequent decisions, and intention
interference. These features, although being quantum in their description,
at the same time, have natural and transparent interpretations in the
simple everyday language and are applicable to the events of the real life.
To stress the applicability of the approach to the decision making of 
human beings, we have provided a number of simple illustrative examples.

We have demonstrated the applicability of our approach to the cases when
the Savage sure-thing principle is violated, resulting in the disjunction
effect. Interference of intentions, arising in decision making under
uncertainty, possesses specific features caused by aversion to uncertainty.
The theorem of interference alternation that we have derived connects the
aversion to uncertainty to the appearance of negative interference terms
suppressing the probability of actions. At the same time, the probability
of the decision maker not to act is enhanced by positive interference
terms. This alternating nature of the intention interference under
uncertainty explains the occurrence of the disjunction effect.

We have proposed a calculation of the interference terms, based on
considerations using robust assessment of probabilities, which makes it
possible to predict their influence in a quantitative way. The estimates
are in good agreement with experimental data for the disjunction effect.

The conjunction fallacy, demonstrated by individual decision makers, is 
also explained by the presence of the interference terms. A series of 
experiments are analysed and shown to be in excellent agreement with the 
a priori evaluation of interference effects. The conjunction fallacy is
also shown to be a sufficient condition for the disjunction effect and 
novel experiments testing the combined interplay between the two effects 
are suggested.

The main features of the Quantum Decision Theory can be summarized as 
follows.

\vskip 2mm

(1) Quantum Decision Theory is a general mathematical approach that is 
applicable to arbitrary situations. We do not try to adjust QDT to 
fit particular cases, but the same theory is used throughout the paper 
to treat quite different effects.

\vskip 2mm

(2) Mathematically, QDT is based on the theory of Hilbert spaces and 
techniques that have been developed in quantum theory. However the use
of these techniques serves only as a convenient formal tool, implying 
no quantum nature of decision makers. 

\vskip 2mm

(3) Each decision maker possesses his/her own strategic state of mind, 
characterizing this decision maker as a separate individual.  

\vskip 2mm

(4) The QDT developed here allows us to characterize not a single unusual, quantum-like, 
property of the decision making process, but several of these 
characteristics, including entangled decisions and the interference between 
intentions.

\vskip 2mm

(5) Aversion with respect to uncertainty is an important feeling regulating 
decision making. We formulate this general and ubiquitous feeling under the 
uncertainty-aversion principle, connecting it to the signs of the 
alternating interference terms. 

\vskip 2mm

(6) We prove the theorem on interference alternation, which shows that
the interference between several intentions, arising under uncertainty,
consists of several terms alternating in sign, some being positive and
some being negative. These  terms are the source of the different
paradoxes and logical fallacies presented by humans making decisions in
uncertain contexts.

\vskip 2mm

(7) Uncertainty aversion and interference alternation, combined together, 
are the key factors that suppress the probability of acting and, at the
same time, enhance the probability of remaining passive, in the case of 
uncertainty.

\vskip 2mm

(8) We demonstrate that it is not simply the interference between
intentions as such, but specifically the interference alternation,
together with the uncertainty aversion, which is responsible for the
violation of Savage's sure-thing principle at the origin of the
disjunction effect.

\vskip 2mm

(9) The conjunction fallacy is another effect that is caused by the
interference of intentions, together with the uncertainty-aversion
principle. Without the latter, the conjunction fallacy cannot be
explained. 

\vskip 2mm

(10) The conjunction fallacy is shown to be a sufficient condition
for the disjunction effect to occur, exhibiting a deep link between the
two effects.

\vskip 2mm

(11) The general ``interference-quarter law" is formulated, which provides
a quantitative prediction for the amplitude of the interference terms,
and thus of the quantitative level by which the sure-thing principle is
violated.

\vskip 2mm

(12) Detailed quantitative comparisons with experiments, documenting the
disjunction effect and the conjunction fallacy, confirm the validity of
the derived laws.

\vskip 5mm

{\bf Acknowledgements}. We are very grateful to E.P. Yukalova for 
many discussions and useful advice. We also acknowledge helpful 
correspondence with P.A. Benioff, J.R. Busemeyer, and Y. Malevergne. We
appreciate the highly constructive advices of the referees, which helped us 
to improve the presentation of our approach.

\newpage

\noindent
{\Large {\bf References}}

\vskip 5mm

\parindent=0pt

Allais, M. (1953), Le comportement de l'homme rationnel devant le risque:
critique des postulats et axiomes de l'ecole Am\'ericaine, 
{\it Econometrica} 21, 503-546.

\vskip 2mm
Al-Najjar, N.I. and Weinstein, J. (2009), The ambiguity aversion 
literature: a critical assessment, Kellogg School of Management Working 
Paper, Northwestern University. 

\vskip 2mm
Atmanspacher, H. (2003), Mind and matter as asymptotically disjoint
inequivalent representations with broken time-reversal symmetry,
{\it Biosystems} 68, 19-30.

\vskip 2mm
Atmanspacher, H., R\"omer, H. and Walach, H. (2002), Weak quantum theory:
complementarity and entanglement in physics and beyond, {\it Foundation
of Physics} 22, 379-406.

\vskip 2mm
Bather, J. (2000), {\it Decision Theory}, Wiley, Chichester.

\vskip 2mm
Barkan, R., Danziger, S., Ben-Bashat, G. and Busemeyer, J.R. (2005), 
Framing reference points: the effect of integration and segregation on 
dynamic inconsistency, {\it Journal of Behavioral Decision Making}
18, 213-226.

\vskip 2mm
Baron, J. (1998), {\it Judgement Misguided: Intuition and Error in Public
Decision-Making}, Oxford University, Oxford.

\vskip 2mm
Bechara, A., Damasio, H. and Damasio, A. (2000), Emotion, decision making
and the orbitofrontal cortex, {\it Cerebral Cortex} 10, 295-307.

\vskip 2mm
Beck, F. and Eccles, J. (1992), Quantum aspects of brain activity and the
role of consciousness, {\it Proceedings of National Academy of Sciences
of USA} 89, 11357-11361.

\vskip 2mm
Bell, J.S. (1964), On the Einstein-Podolsky-Rosen paradox, {\it Physics} 1,
195-200.

\vskip 2mm
Benioff, P.A. (1972), Decision procedures in quantum mechanics, {\it Journal
of Mathematical Physics} 13, 909-915.

\vskip 2mm
Benjamin, S.C. and Hayden, P.M. (2001), Multi-player quantum games, 
{\it Physical Review} A 64, 030301.

\vskip 2mm
Berger, J.O. (1985), {\it Statistical Decision Theory and Bayesian Analysis}, 
Springer, New York.

\vskip 2mm
Bohr, N. (1929), Wirkungsquantum und Naturbeschreibung, 
{\it Naturwissenschaft} 17, 483-486.

\vskip 2mm
Bohr, N. (1933), Light and life, {\it Nature} 131, 421-423, 457-459.

\vskip 2mm
Bohr, N. (1937), Kausalit\"at und Komplemetarit\"at., 
{\it Erkenntnissenscahft} 6, 293-303.

\vskip 2mm
Bohr, N. (1961), {\it La Physique Atomique et la Connaissance Humaine},
Gontier, Gen\`eve.

\vskip 2mm
Buchanan, J.T. (1982), {\it Discrete and Dynamic Decision Analysis},
Wiley, Chichester.

\vskip 2mm
Busemeyer, J.R., Wang, Z. and Townsend, J.T. (2006), Quantum dynamics
of human decision-making, {\it Journal of Mathematical Psychology} 50, 
220-241.

\vskip 2mm
Camerer, C. F., Loewenstein, G. and Rabin, R., eds. (2003)
{\it Advances in Behavioral Economics}, Princeton University, Princeton.

\vskip 2mm
Chalmers, D. (1996), {\it The Conscious Mind}, Oxford University, Oxford.

\vskip 2mm
Charness, G.B., Levin, D. and Karni, E. (2008), On the conjunction fallacy 
in probability judgement: new experimental evidence, Department of Economics 
Working Paper, UCSB.

\vskip 2mm
Cohen, M. and Tallon, J.M. (2000), D\'ecision dans le risque et l'incertain:
l'apport des mod\`eles non additifs, {\it Revue d'Economie Politique} 110,
631-681.

\vskip 2mm
Croson, R.T.A. (1999), The disjunction effect and reason-based choice
in games, {\it Organizational Behavior and Human Decision Processes} 80, 
118-133.

\vskip 2mm
Dickhaut, J., McCabe, K., Nagode, J.C., Rustichini, A., Smith, K. and
Pardo, J.V. (2003), The impact of the certainty context on the process
of choice, {\it Proceedings of National Academy of Sciences of USA} 100, 
3536-3541.

\vskip 2mm
Dieudonn\'e, J. (2006), {\it Foundations of Modern Analysis}, 
Hesperides, London.

\vskip 2mm
Dirac, P.A.M. (1958), {\it The Principles of Quantum Mechanics},
Clarendon, Oxford.

\vskip 2mm
Du, J., Li, H., Xu, X., Shi, M., Wu, J., Zhou, X. and Han, R. (2002),
Experimental realization of quantum games on a quantum computer, 
{\it Physical Review Letters} 88, 137902.

\vskip 2mm
Du, J., Xu, X., Li, H., Zhou, X. and Han, R. (2001), Entanglement
playing a dominating role in quantum games, {\it Physics Letters} 
A 289, 9-15.

\vskip 2mm
Einstein, A., Podolsky, B. and Rosen, N. (1935), Can quantum-mechanical 
description of physical reality be considered complete? 
{\it Physical Review} 47, 777-780.

\vskip 2mm
Eisert, J. and Wilkens, M. (2000), Quantum games, {\it Journal of Modern 
Optics} 47, 2543-2556.

\vskip 2mm
Enk, van S.J. and Pike, R. (2002), Classical rules in quantum games.
{\it Physical Review} A 66, 024306.

\vskip 2mm
Epstein, L.G. (1999), A definition of uncertainty aversion,
{\it The Review of Economic Studies} 66, 579-608.

\vskip 2mm
Feller, W. (1970), {\it Introduction to Probability Theory and Its
Applications}, Wiley, New York.

\vskip 2mm
Fox, C., Rogers, B. and Tversky, A. (1996), Option traders exhibit
subadditive decision weights, {\it Journal of Risk and  Uncertainty} 
13, 5-17.

\vskip 2mm
Frederick, S., Loewenstein, G. and O'Donoghue, T. (2002), Time discounting
and time preference: a critical review, {\it Journal of Economic Literature}
40, 351-401.

\vskip 2mm
French, S. and Insua, D.R. (2000), {\it Statistical Decision Theory}, 
Arnold, London.

\vskip 2mm
Fr\"olich, H. (1968), Bose condensation of strongly excited longitudinal
electric modes, {\it Physics Letters} A 26, 402-403.

\vskip 2mm
Gilboa, I. (1987), Expected utility with purely subjective 
non-additive probabilities, {\it Journal of Mathematical Economics} 16,
65-88.

\vskip 2mm
Gilboa, I. and Schmeidler, D. (1989), Maxmin expected utility with 
non-unique prior, {\it Journal of Mathematical Economics} 18, 141-153.

\vskip 2mm
Goldenberg, L., Vaidman, L. and Wiesner, S. (1999), Quantum gambling. 
{\it Physical Review Letters} 82, 3356-3359.

\vskip 2mm
Hagan, S., Hameroff, S.R. and Tuszynski, J.A. (2002), Quantum 
computation in brain microtubules: decoherence and biological 
feasibility, {\it Physical Review} E 65, 061901.

\vskip 2mm
Hastings, N.A. and Mello, J.M. (1978), {\it Decision Networks}, 
Wiley, Chichester.

\vskip 2mm
Iqbal, A. and Toor, A.H. (2001), Evolutionally stable strategies in
quantum games, {\it Physics Letters} A 280, 249-256.

\vskip 2mm
Johnson, N.F. (2001), Playing a quantum game with a corrupted source,
{\it Physical Review} A 63, 020302.

\vskip 2mm
Kaplan, S. and Garrick, B.J. (1981), On the quantitative definition of 
risk, {\it Risk Analysis} 1, 11-27.

\vskip 2mm
Koechlin, E. and Hyafil, A. (2007), Anterior prefrontal function and 
the limits of human decision-making, {\it Science} 318, 594-598.

\vskip 2mm
K\"uhberger, A., Komunska D. and Perner, J. (2001), The disjunction
effect: does it exist for two-step gambles? {\it Organizational 
Behavior and Human Decision Processes} 85, 250-264.

\vskip 2mm
Lambdin, C. and Burdsal, C. (2007), The disjunction effect reexamined:
relevant methodological issues and the fallacy of unspecified 
percentage comparisons, {\it Organizational Behavior and Human Decision 
Processes} 103, 268-276.

\vskip 2mm
Lee, C.F. and Johnson, N.F. (2003), Efficiency and formalism of 
quantum games, {\it Physical Review} A 67, 022311.

\vskip 2mm
Legrenzi, P., Girotto, V. and Johnson-Laird, P.N. (1993), Focusing 
in reasoning and decision making, {\it Cognition} 49, 36-66.

\vskip 2mm
Li, C.F., Zhang, Y.S., Huang, Y.F. and Guo, G.C. (2001), Quantum
strategies of quantum measurements, {\it Physics Letters} A 280, 
257-260.

\vskip 2mm
Li, S., Taplin, J.E. and Zhang, Y. (2007), The equate-to-differentiate 
way of seeing the prisoner's dilemma, {\it Information Sciences} 177, 
1395-1412.

\vskip 2mm
Lindgren, B.W. (1971), {\it Elements of Decision Theory}, Macmillan, 
New York.

\vskip 2mm
Lockwood, M. (1989), {\it Mind, Brain and the Quantum}, Basil 
Blackwell, Oxford.

\vskip 2mm
Machina, M.J. (2008), Non-expected utility theory, in {\it The New 
Palgrave Dictionary of Economics}, Durlauf, S.N. and Blume, L.E. (eds.) 
Macmillan, Basingstoke.

\vskip 2mm
Marshall, K.T. and Oliver, R.M. (1995), {\it Decision Making and 
Forecasting}, McGraw-Hill, New York.

\vskip 2mm
Mendelson, E. (1965), {\it Introduction to Mathematical Logic},
Van Nostrand, Princeton.

\vskip 2mm
Meyer, D. (1999), Quantum strategies, {\it Physical Review Letters} 
82, 1052-1055.

\vskip 2mm
Montesano, A. (2008), Effects of uncertainty aversion on the call 
option market, {\it Theory and Decision} 65, 97-123.

\vskip 2mm
Morier, D.M. and Borgida, E. (1984), The conjunction fallacy: 
a task-specific phenomenon? {\it Personality Social Psychology 
Bulletin} 10, 243-253.

\vskip 2mm
Neumann, von J. and Morgenstern, O. (1944), {\it Theory of Games 
and Economic Behavior}, Princeton University, Princeton.

\vskip 2mm
Neumann, von J. (1955), {\it Mathematical Foundations of Quantum 
Mechanics}, Princeton University, Princeton.

\vskip 2mm
Penrose, R. (1989), {\it The Emperor's New Mind}, Oxford University, 
Oxford.

\vskip 2mm
Pessa, E. and Vitiello, G. (2003), Quantum noise, entanglement and chaos
in the quantum field theory of mind-brain states, {\it Mind and Matter} 
1, 59-79.

\vskip 2mm
Primas, H. (2003), Time-entanglement between mind and matter, {\it Mind
and Matter} 1, 81-119.

\vskip 2mm
Quiggin, J. (1982), A theory of anticipated utility, {\it Journal of 
Economic Behavior and Organization} 3, 323-343.

\vskip 2mm
Raiffa, H. and Schlaifer, R. (2000), {\it Applied Statistical 
Decision Theory}, Wiley, New York.

\vskip 2mm
Read, D., Loewenstein, G.L. and Rabin, M. (1999), Choice bracketing, 
{\it Journal of Risk and Uncertainty} 19, 171-197.

\vskip 2mm
Rivett, P. (1980), {\it Model Building for Decision Analysis}, 
Wiley, Chichester.

\vskip 2mm
Rottenstreich Y. and Tversky, A. (1997), Unpacking, repacking and
anchoring: advances in support theory, {\it Psychological Review} 104, 
406-415.

\vskip 2mm
Safra, Z. and Segal, U. (2008), Calibration results for non-expected 
utility theories, {\it Econometrica} 76, 1143-1166.

\vskip 2mm
Satinover, J. (2001), {\it The Quantum Brain}, Wiley, New York.

\vskip 2mm
Savage, L.J. (1954), {\it The Foundations of Statistics}, Wiley, 
New York.

\vskip 2mm
Schmeidler, D. (1989), Subjective probability and expected utility 
without additivity, {\it Econometrica} 57, 571-587.

\vskip 2mm
Shafir, E.B., Smith, E.E. and Osherson, D.N. (1990), Typicality and 
reasoning fallacies, {\it Memory and Cognition} 18, 229-239.

\vskip 2mm
Shafir, E. and Tversky, A. (1992), Thinking through uncertainty:
nonconsequential reasoning and choice, {\it Cognitive Psychology}
24, 449-474.

\vskip 2mm
Shafir, E., Simonson, I. and Tversky, A. (1993), Reason-based choice,
{\it Cognition} 49, 11-36.

\vskip 2mm
Shafir, E. (1994), Uncertainty and the difficulty of thinking through
disjunctions, {\it Cognition} 50, 403-430.

\vskip 2mm
Sheremeta, R.M. and Zhang, J. (2009), Can groups solve the problem of 
over-building in contests? Department of Economics Working Paper, 
McMaster University.

\vskip 2mm
Shor, P. (1997), Polynomial-type algorithms for prime factorization
and discrete logarithms on a quantum computer, {\it SIAM Journal of 
Scientific and Statistical Computing} 26, 1484-1494.

\vskip 2mm
Simon, H.A. (1955), A behavioral model of rational choice, {\it Quarterly 
Journal of Economics} 69, 99-118.

\vskip 2mm
Sornette, D., Davis, A.B., Ide, K., Vixie, K.R., Pisarenko, V. and
Kamm, J.R. (2007), Algorithm for model validation: theory and 
applications, {\it Proceedings of National Academy of Sciences of USA}
104, 6562-6567.

\vskip 2mm
Stapp, H.P. (1993), {\it Mind, Matter, and Quantum Mechanics}, 
Springer, Berlin.

\vskip 2mm
Stapp, H.P. (1999), Attention, intention, and will in quantum physics, 
{\it Journal of Consciousness Studies} 6, 143-164.

\vskip 2mm
Stuart, C.I.J., Takahashi, Y. and Umezawa, H. (1978), On the stability 
and non-local properties of memory, {\it Journal of Theoretical Biology} 
71, 605-618.

\vskip 2mm
Stuart, C.I.J., Takahashi, Y. and Umezawa, H. (1979), Mixed system 
brain dynamics: neural memory as a macroscopic ordered state, 
{\it Foundations of Physics} 9, 301-327.

\vskip 2mm
Tegmark, M. (2000), Importance of quantum decoherence in brain 
processes, {\it Physical Review} E 61, 4194-4205.

\vskip 2mm
Tentori, K., Bonini, N. and Osherson, D. (2004), The conjunction 
fallacy: a misunderstanding about conjunction? {\it Cognitive Science}
28, 467-477.

\vskip 2mm
Tversky, A. and Kahneman, D. (1973), Availability: a heuristic for 
judging frequency and probability, {\it Cognitive Psychology} 5, 
207-232.

\vskip 2mm
Tversky, A. and Kahneman, D. (1980), Judgements of and by 
representativeness, in {\it Judgements Under Uncertainty: Heuristics 
and Biases}, Kahneman, D., Slovic, P. and Tversky, A. (eds.), 
Cambridge University, New York, p. 84-98.

\vskip 2mm
Tversky, A. and Kahneman, D. (1983), Extensional versus intuitive 
reasoning: the conjunction fallacy in probability judgement,
{\it Psychological Review} 90, 293-315.

\vskip 2mm
Tversky, A. and Shafir, E. (1992), The disjunction effect in choice 
under uncertainty, {\it Psychological Science} 3, 305-309.

\vskip 2mm
Tversky, A. and Koehler, D. (1994), Support theory: a nonexistential
representation of subjective probability, {\it Psychological Review} 
101, 547-567.

\vskip 2mm
Vitiello, G. (1995), Dissipation and memory capacity in the quantum
brain model, {\it International Journal of Modern Physics} B 9, 
973-989.

\vskip 2mm
Wells, G.L. (1985), The conjunction error and the representativeness
heuristic, {\it Social Cognition} 3, 266-279.

\vskip 2mm
Weirich, P. (2001), {\it Decision Space}, Cambridge University, 
Cambridge.

\vskip 2mm
White, D.I. (1976), {\it Fundamentals of Decision Theory}, Elsevier, 
New York.

\vskip 2mm
Yates, J.F. and Carlson, B.W. (1986), Conjunction errors: evidence for
multiple judgement procedures, including signed summation, 
{\it Organizational Behavior and Human Decision Processes} 37, 230-253.

\vskip 2mm
Yukalov, V.I. (1975), Causality problem in quantum physics, 
{\it Philosophical Sciences} 18, 145-147.

\vskip 2mm
Yukalov, V.I. (2003), Entanglement measure for composite systems,
{\it Physical Review Letters} 90, 167905.

\vskip 2mm
Yukalov, V.I. (2003), Quantifying entanglement production of quantum
operations, {\it Physical Review} A 68, 022109.

\vskip 2mm
Yukalov, V.I. (2003), Evolutional entanglement in nonequilibrium
processes, {\it Modern Physics Letters} B 17, 95-103.

\vskip 2mm
Yukalov, V.I. and Sornette, D. (2008), Quantum decision theory as 
quantum theory of measurement, {\it Physics Letters} A 372, 6867-6871.

\vskip 2mm
Yukalov, V.I. and Sornette, D. (2009), Scheme of thinking quantum systems,
{\it Laser Physics Letters} 6, 833-839.

\vskip 2mm
Yukalov, V.I. and Sornette, D. (2009), Physics of risk and uncertainty 
in quantum decision making, {\it European Physical Journal} B 71, 533-548.

\vskip 2mm
Zeckhauser, R. (2006), Investing in the unknown and unknowable,
{\it Capitalism and Society} 1, 1-39.

\newpage

\parindent=0pt

\begin{tabular}{|c|c|c|c|c|}  \hline
  & characteristics & $p(AX)$ & $p(AX_1)$ & $q(AX)$ \\ \hline
$A$   & bank teller      &   0.241 & 0.401 & -0.259 \\
$X_1$ & feminist        &         &       &       \\ \hline
$A$   & bird watcher    & 0.173   & 0.274 & -0.327 \\
$X_1$ & truck driver    &         &       &       \\ \hline
$A$   & bicycle racer   & 0.160   & 0.226 & -0.340 \\
$X_1$ & nurse           &         &       &        \\ \hline
$A$   & drum player     & 0.266   & 0.367 & -0.234 \\
$X_1$ & professor       &         &       &        \\ \hline
$A$   & boxer           & 0.202   & 0.269 & -0.298 \\
$X_1$ & chef            &         &       &         \\ \hline
$A$   & volleyboller    & 0.194   & 0.282 & -0.306 \\
$X_1$ & engineer        &         &       &        \\ \hline
$A$   & librarian       & 0.152   & 0.377 & -0.348 \\
$X_1$ & aerobic trainer &         &       &        \\ \hline
$A$   & hair dresser    & 0.188   & 0.252 & -0.312  \\
$X_1$ & writer          &         &       &     \\ \hline
$A$   & floriculturist  & 0.310   & 0.471 & -0.190 \\
$X_1$ & state worker    &         &       &    \\ \hline
$A$   & bus driver      & 0.172   & 0.314 & -0.328 \\
$X_1$ & painter         &         &       &    \\ \hline
$A$   & knitter         & 0.315   & 0.580 & -0.185 \\
$X_1$ & correspondent   &         &       &   \\ \hline
$A$   & construction worker & 0.131 & 0.249 & -0.369 \\
$X_1$ & labor-union president &    &      &    \\ \hline
$A$   & flute player    & 0.180   & 0.339 & -0.320 \\
$X_1$ & car mechanic    &         &       &    \\ \hline
$A$   & student         & 0.392   & 0.439 & -0.108 \\
$X_1$ & fashion-monger   &         &       &       \\ \hline
  & average         & 0.220   & 0.346 & -0.280 \\ \hline
\end{tabular}

\vskip 3mm

\parindent=0pt
{\bf Table 1.} Conjunction fallacy and related interference terms caused
by the decision under uncertainty. The average interference term is in
good agreement with the interference-quarter law. The empirical data are 
taken from Shafir et al. (1990).

\end{document}